\documentclass[a4paper,11pt,amsmath,amssymb,amsfonts,superscriptaddress,notitlepage,reprint]{revtex4-1}

\usepackage{graphicx} % manages external pictures
\usepackage{hyperref}
\usepackage{color} % basic color, strikethru
\usepackage{ulem} % strikethru command
\usepackage{lmodern} % enhanced version of Computer Modern fonts, available as Type 1
\usepackage{textcomp} %provides extra symbols like \textcelsius, etc.
\usepackage[T1]{fontenc} % utf8 encoding

\newcommand*{\ket}[1]{|{#1}\rangle}

\newcommand*{\mean}[1]{\mathinner{\langle{#1}\rangle}}
\newcommand*{\braket}[2]{\mathinner{\langle{#1}|{#2}\rangle}}

\newcommand*{\twobytwo}[4]{\begin{pmatrix} #1 & #2 \\ #3 & #4 \end{pmatrix}}
\newcommand*{\equalsup}[2]{\frac{1}{\sqrt{2}}\left(\ket{#1}+\ket{#2}\right)}

\newcommand*{\trans}[2]{\ket{#1}\leftrightarrow\ket{#2}}
\def\be{\begin{equation}}
\def\ee{\end{equation}}
\def\bes{\begin{equation*}}
\def\ees{\end{equation*}}

\newcommand*{\fref}[1]{Fig.~\ref{#1}}
\newcommand*{\mywidth}{8.6 cm}
\newcommand*{\pA}{(\textbf{A})~}
\newcommand*{\pB}{(\textbf{B})~}
\newcommand*{\pC}{(\textbf{C})~}
\newcommand*{\pD}{(\textbf{D})~}

\newcommand*{\nph}{n}
\newcommand*{\dtn}{\Delta}
\newcommand*{\coup}{g}
\newcommand*{\opcoup}{\coup}
\newcommand*{\eigens}{\ket{\Psi_\nph^\pm}}

\newcommand*{\wone}{\omega_1}
\newcommand*{\wtwo}{\omega_2}
\newcommand*{\wthr}{\omega_3}
\newcommand*{\wmac}{\omega_d}
\newcommand*{\plen}{\tau}
\newcommand*{\mixngl}{\Theta_{\nph}}
\newcommand*{\eigpl}{\ket{\Psi_\nph^+}}
\newcommand*{\eigmn}{\ket{\Psi_\nph^-}}
\newcommand*{\dphi}{\delta\varphi}

\begin{document}

\title{Measurement of a Vacuum-Induced Geometric Phase}
\author{S. Gasparinetti}
\email{gasimone@phys.ethz.ch}
\affiliation{Department of Physics, ETH Zurich, CH-8093 Zurich, Switzerland}
\author{S. Berger}
\affiliation{Department of Physics, ETH Zurich, CH-8093 Zurich, Switzerland}
\author{A. A. Abdumalikov}
\affiliation{Department of Physics, ETH Zurich, CH-8093 Zurich, Switzerland}
\author{M. Pechal}
\affiliation{Department of Physics, ETH Zurich, CH-8093 Zurich, Switzerland}
\author{S. Filipp}
\affiliation{IBM T.J. Watson Research Center, Yorktown Heights, NY 10598, USA}
\author{A. Wallraff}
\affiliation{Department of Physics, ETH Zurich, CH-8093 Zurich, Switzerland}

\begin{abstract}
Berry's geometric phase naturally appears when a quantum system is driven by an external field whose parameters are slowly and cyclically changed. A variation in the coupling between the system and the external field can also give rise to a geometric phase, even when the field is in the vacuum state or any other Fock state. Here we demonstrate the appearance of a vacuum-induced Berry phase in an artificial atom, a superconducting transmon, interacting with a single mode of a microwave cavity. As we vary the phase of the interaction, the artificial atom acquires a geometric phase determined by the path traced out in the combined Hilbert space of the atom and the quantum field. Our ability to control this phase opens new possibilities for the geometric manipulation of atom-cavity systems also in the context of quantum information processing.
\end{abstract}

\date{\today}

\maketitle

%Intro
Geometric phases are at the heart of many phenomena in solid-state physics \cite{Xiao2010}, from the quantum Hall effect \cite{Thouless1982} to topological phases \cite{Hasan2010,Qi2011}, and may provide a resource for quantum computation \cite{Zanardi1999,Sjoqvist2008}.
As a quantum system is steered in its state space by controlled interaction with an external field, the trajectory it describes can be associated with a geometric phase \cite{Berry1984}.
While the external field is typically treated as classical, at low excitation numbers its quantization is expected to produce novel geometric effects \cite{Fuentes-Guridi2002}.
Here we experimentally demonstrate that a geometric phase of Berry's type \cite{Berry1984} can be induced by a variable coupling between the system and a quantized field, using a superconducting circuit.
This phase is nonvanishing even when the quantized field is in the vacuum state, a result with no semiclassical analogue. It has been referred to as the \textit{vacuum}-induced Berry phase \cite{Fuentes-Guridi2002} and its existence and observability have been subject of a theoretical debate \cite{Fuentes-Guridi2002,Liu2011a,Larson2012,Wang2015}. According to Larson \cite{Larson2012}, it is an artifact of the rotating-wave approximation. However, later work by Wang \textit{et al.}~\cite{Wang2015} shows that a vacuum-induced Berry phase is always associated to the Rabi model, regardless of whether the rotating-wave approximation is used or not. No evidence of this phase has so far been observed, possibly due to the difficulty in engineering the relevant interaction, while superconducting circuits have already been used to study geometric phases \cite{Falci2000, Leek2007, Mottonen2008, Pechal2012,Abdumalikov2013}, their susceptibility to noise \cite{Berger2013}, and their relation to topological effects \cite{Roushan2014,Schroer2014}.

\bigskip

In previous measurements of the Berry phase \cite{Jones2000,Leek2007}, a transition between two quantum states $\ket{g}$ and $\ket{f}$ was driven by a coherent field of amplitude $\alpha$, detuning $\Delta$ and phase $\varphi$ [Fig.~1(A)]. In a frame rotating at the drive frequency, the corresponding dynamics is that of a spin-$1/2$ particle interacting with an effective magnetic field $\vec B = (2\opcoup \alpha \cos \varphi, 2\opcoup \alpha \sin \varphi, \Delta)$, where $\opcoup$ is the dipole strength of the transition.
An adiabatic variation of $\varphi$ causes $\vec B$ to precess about the $\hat z$ axis;
the corresponding path traced out by the spin particle in its Hilbert space can be obtained by projecting the vector $\vec B$ onto the Bloch sphere [Fig.~1(B)].
The spin particle, initially in its ground state, acquires a geometric phase $\gamma=\Omega/2$, where $\Omega$
is the solid angle subtended by the circular path \cite{Berry1984}.

As noticed by Fuentes-Guridi \textit{et al.}~\cite{Fuentes-Guridi2002}, the model of Fig.~1(A) is a semiclassical one: it ignores the quantization of the applied field and neglects the effect of vacuum fluctuations on the Berry phase.
By contrast, a fully quantized version of the problem is captured by the Hamiltonian
\be
\hat H=\frac{\Delta}{2}\hat\sigma_z + \coup\left(\hat \sigma_+ \hat a e^{-i\varphi} + \hat \sigma_-\hat a^\dagger e^{i\varphi}\right) \ , \label{eq:Hquant}
\ee
where $\hat\sigma_z$ is a Pauli matrix acting on the Hilbert space $\{\ket{g},\ket{f}\}$ of the two-level system, $\Delta$ is the detuning of the quantized field, $\coup$ is the coupling, and $\hat a$, $\hat a^\dagger$, $\hat \sigma_-$, and $\hat \sigma_+$ are the annihilation and creation operators of the quantized field and the two-level system, respectively.
The Hamiltonian \eqref{eq:Hquant} describes a Jaynes-Cummings-type interaction
with a variable phase $\varphi$ and
gives rise to a finite Berry phase also in the limit of vanishing photon number \cite{Fuentes-Guridi2002}.

In our experiment, we realize a tunable coupling between a cavity mode and two levels $\ket{g}$ and $\ket{f}$ of a superconducting artificial atom by applying a coherent microwave signal \cite{Bose2003, Liu2010, Pechal2014, Zeytinoglu2015}, as schematically shown in Fig.~1(C) and detailed in the following.
A slow modulation of the coupling phase realizes a geometric manipulation which is the quantum analogue of the semiclassical evolution depicted in Fig.~1(A,B). To understand its effects, consider the eigenstates of the Hamiltonian \eqref{eq:Hquant}.
The ground state $\ket{g,0}$ is not coupled to any other state; as such, it acquires no geometric phase. The other eigenstates are coupled in pairs $\ket{\Psi_{\nph}^\pm}$ having support in the subspace $\{ \ket{f,\nph}, \ket{g,\nph+1} \}$, with $\nph$ denoting the photon number in the cavity. As $\varphi$ is adiabatically steered, each subspace
undergoes a different evolution, shown in Fig.~1(D) for the first few photon numbers. 
The geometric phase accumulated by the states $\eigens$ is given by \cite{Fuentes-Guridi2002}
\be
\gamma_\nph^\pm = \pi\left[1 \mp \frac{\dtn}{\sqrt{\dtn^2+4 \coup^2 (\nph+1)}} \right] \ .
\label{eq:gammadtn}
\ee
A comparison to Fig.~1(B) highlights two key features of the quantized model: (i) for a given coupling $\coup$ and detuning $\Delta$, only a discrete set of paths are admissible, corresponding to integer values of $\nph$, and (ii) a finite solid angle is enclosed even when $\nph=0$, corresponding to a vacuum-induced Berry phase.

\begin{figure}
\centering
	\includegraphics{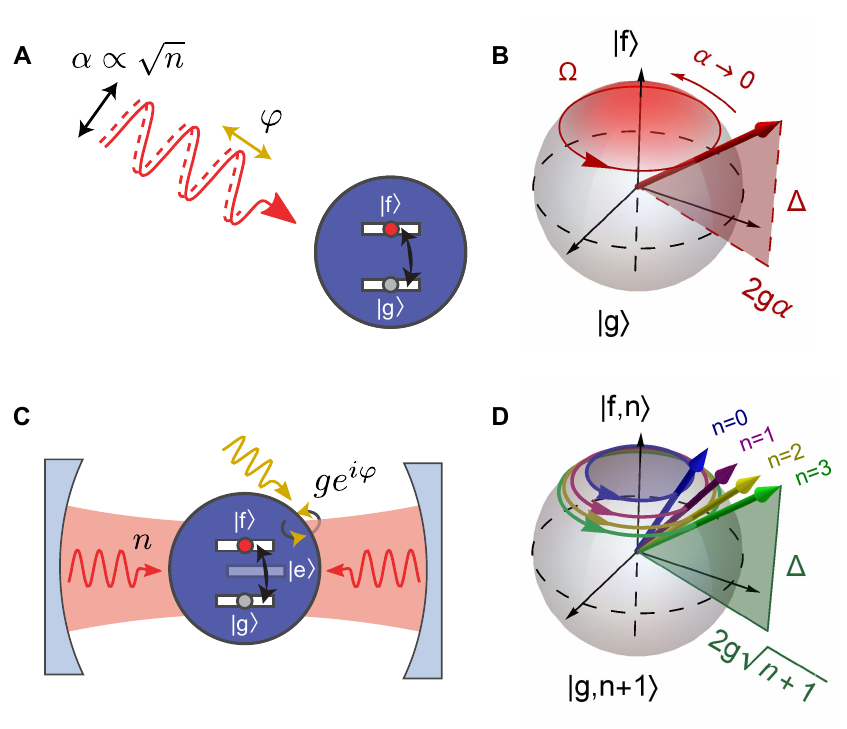}
	\caption{\textbf{Berry phase induced by a quantized field.}
\pA An atomic transition between two states $\ket{g}$ and $\ket{f}$ is driven by a coherent tone of amplitude $\alpha$, phase $\varphi$, and detuning $\Delta$. The phase $\varphi$ is slowly varied between zero and $2\pi$.
\pB In a frame rotating at the drive frequency, the drive acts as an effective magnetic field (red, thick arrow) precessing about the $\hat z$ axis. In the adiabatic limit, the Bloch vector stays aligned with the field and describes a circular path on the Bloch sphere spanned by the atomic basis states $\ket{g}$ and $\ket{f}$.
The acquired geometric phase equals half the solid angle $\Omega$ subtended by the path.
\pC By placing the atom in a cavity, the atom interacts with a quantized field. The interaction between the atom and the field is controlled by a microwave-activated coupling, which is mediated by an intermediate state $\ket{e}$ and is tunable in amplitude ($\coup$) and phase ($\varphi$).
\pD
Admissible paths on the Bloch sphere for different numbers of photons $n$ in the cavity. For each $n$, the Bloch sphere is that spanned by the basis states $\ket{g,n+1}$ and $\ket{f,n}$ of the combined atom-cavity system.
}
	\label{fig:impress}
\end{figure}

\bigskip

\begin{figure}
\centering
	\includegraphics[width=\mywidth]{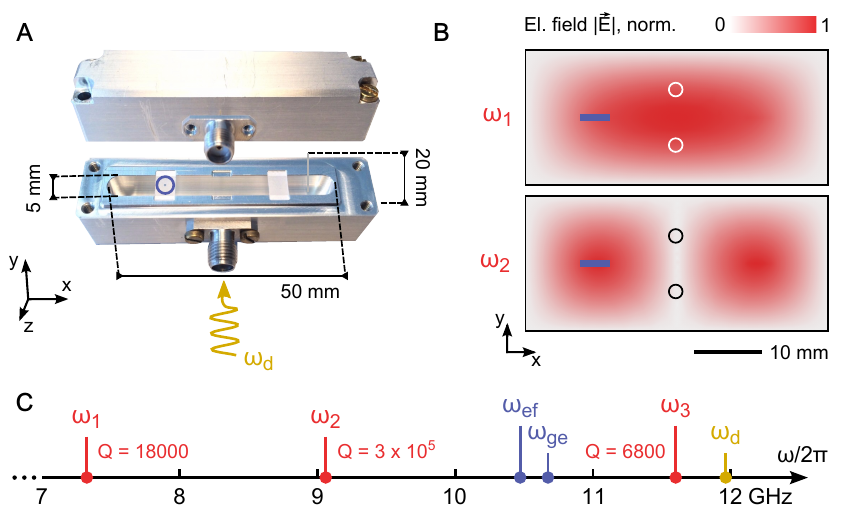}
	\caption{\textbf{Transmon in a 3D cavity with mode-selective coupling ports.}
\pA Edited photograph of the 3D cavity used in the experiment. Two sapphire chips are placed inside the cavity. A transmon is patterned on the left chip (blue circle).
\pB Cross-section of the electric-field magnitude $|\vec{E}|$ for the two lowest-frequency modes of the cavity in (A), as obtained from a finite-element simulation. The chip used in the experiment is highlighted in blue and the cavity ports are indicated as circles (not to-scale).
\pC Diagram of the relevant frequencies for the experiment: first three cavity modes, $\wone$, $\wtwo$ and $\wthr$ (red), first two transitions of the transmon, $\omega_{ge}$ and $\omega_{ef}$ (blue), and microwave-activated coupling between states $\ket{f0}$ and $\ket{g1}$, $\wmac$ (yellow).
}
	\label{fig:intro}
\end{figure}

%Experimental set-up
Our set-up consists of a superconducting transmon-type structure embedded in a three-dimensional microwave cavity \cite{Paik2011, Abdumalikov2013}. The cavity, shown in Fig.~2(A), is made of aluminum and has inner dimensions $5 \times 20 \times 50~\rm{mm}$. The distribution of the electric field for the first two modes of the cavity is shown in Fig.~2(B). The position of the coupling ports is such that the first mode is overcoupled while the second mode is strongly undercoupled. The modes have resonant frequencies $\wone/2\pi=7.828~\rm{GHz}$ and $\wtwo/2\pi=9.041~\rm{GHz}$ and quality factors $Q_1=18000$ and $Q_2 \approx 3\times10^5$. The third mode of the cavity has frequency $\wthr/2\pi=11.432~\rm{GHz}$.
The transmon consists of two $200\times300~\mu\rm{m}$ Al pads separated by
$160~\mu\rm{m}$ and connected by a Josephson junction of Josephson energy $E_J/h\approx35~\rm{GHz}$ \cite{Abdumalikov2013}.
The first two transition frequencies of the transmon are $\omega_{ge}/2\pi=10.651~\rm{GHz}$ and $\omega_{ef}/2\pi=10.217~\rm{GHz}$. The decay time of both excited states is $T_1=(4.9 \pm 0.1)~\mu\rm{s}$ and their dephasing time is $T_2^*=(2.0\pm 0.1)~\mu\rm{s}$.
We use the ground and the second excited state of the transmon ($\ket{g}$ and $\ket{f}$, respectively) as the two atomic states and the second mode of the cavity as the quantized field.
To read out the ground, first and second-excited state of the transmon, we measure the state-dependent transmission through the fundamental mode \cite{Bianchetti2010}.
By applying a control field close to the nominal frequency $\wmac=\omega_{ge}+\omega_{ef}-\wtwo$, we induce a microwave-activated coupling between pairs of states $\ket{f,\nph}$ and $\ket{g,\nph+1}$, with amplitude $\coup e^{i\varphi}$ and detuning $\Delta$
\cite{Pechal2014,Zeytinoglu2015,vBp_suppl}.
A diagram of all relevant frequencies for our experiment is shown in Fig.~2(C). When driving the transitions, we compensate for Stark shifts caused by the photonic occupation of the cavity mode \cite{Schuster2007} and the coupling field \cite{Zeytinoglu2015, vBp_suppl}.

\bigskip

\begin{figure}
\centering
	\includegraphics[width=\mywidth]
	{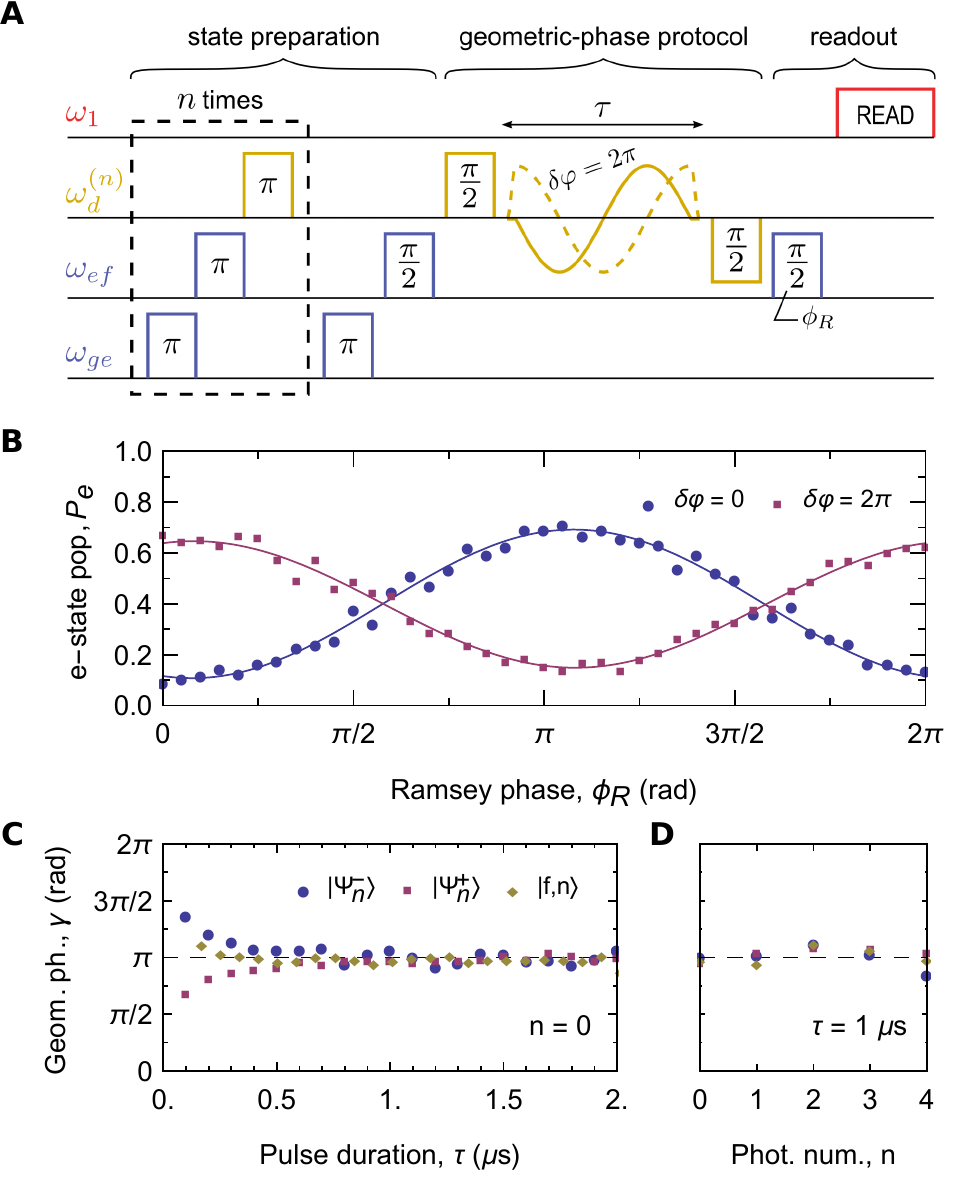}
	\caption{\textbf{Vacuum-induced Berry phase: resonant case.}
	\pA Pulse sequence to detect the geometric phase acquired by the state $\ket{\Psi_{\nph}^-}$ for resonant coupling ($\Delta=0$). The cavity is prepared in an $n$-photon Fock state by repeating the initial sequence $n$ times. The system is prepared in a superposition of $\ket{e,\nph}$ and $\ket{\Psi_{\nph}^-}$. Then the resonant coupling is turned on and its phase is increased by $2\pi$ during a time $\plen$. Finally the relative phase between $\ket{\Psi_{\nph}^-}$ and $\ket{e,\nph}$ is determined by Ramsey interferometry.
	\pB Oscillations observed in the $e$-state population $P_e$ when varying the phase $\phi_R$ of the second Ramsey pulse, with $\tau=420~\rm{ns}$ and $\nph=0$.
	  The measurement described in (A) ($\delta\varphi=2\pi$, red squares) is compared against a reference measurement in which the phase of the coupling is held fixed ($\delta\varphi=0$, blue circles). The phase shift observed in the Ramsey pattern corresponds to an accumulated geometric phase $\gamma_0^- = (3.13 \pm 0.06)$.
  \pC Geometric phase $\gamma$, determined as in (B), versus pulse duration $\plen$. Three different states are prepared: the two eigenstates $\ket{\Psi_0^-}$ (blue circles) and $\ket{\Psi_0^+}$ (red squares), and the state $\ket{f0}$ (yellow diamonds), for which the cavity is initially in the vacuum state.
  \pD Geometric phase accumulated by the states $\ket{\Psi_{\nph}^\pm}$ and $\ket{f,\nph}$, with fixed pulse duration $\plen$ and varying photon number $\nph$.
}
	\label{fig:vBp_Ramsey}
\end{figure}

%Results: resonant case
We first report on measurements done in the resonant case, $\dtn=0$, and with the cavity initially in the vacuum state, $\nph=0$. Our scheme for measuring the geometric phase [\fref{fig:vBp_Ramsey}(A)] relies on the use of $\ket{e}$ as a reference state for Ramsey interferometry. The measured thermal population of $\ket{e}$ is about $1\%$ and is neglected in our analysis. Starting from the ground state $\ket{g0}$ we first
prepare the superposition state $\frac{1}{\sqrt{2}}\left(\ket{f0} + \ket{e0}\right)$ and then apply a resonant coupling pulse to bring the state $\ket{f0}$ into $\ket{\Psi_0^-}\equiv\equalsup{g1}{f0}$.
At this point we again turn on the coupling, choosing its phase so that the effective magnetic field is aligned with the prepared eigenstate $\ket{\Psi_0^-}$ \cite{vBp_suppl}. Then we slowly vary the phase by an amount $\delta\varphi = 2\pi$. A third coupling pulse follows to bring the system back to $\ket{f0}$. The phase carried by $\ket{f0}$, which includes a geometric contribution from the phase manipulation, is finally detected by Ramsey interferometry against the reference state $\ket{e0}$, employing a final $\pi/2$ pulse on the $\trans{e}{f}$ transition with variable phase $\phi_R$.
In order to single out the geometric contribution to the interference phase, we compare patterns obtained with ($\delta\varphi=2\pi$) and without ($\delta\varphi=0$) the phase variation, as the acquired dynamic phase (including Stark shifts) is the same in both cases.
The recorded interference patterns 
clearly oscillate out of phase [\fref{fig:vBp_Ramsey}(B)], with a measured phase shift $\gamma_0^-=(3.13 \pm 0.06)$. This result can be explained by a geometric argument: when $\Delta=0$, the Bloch vector describes a loop on the equator [compare Fig.~1(D)]. The enclosed solid angle is $\Omega=2\pi$, corresponding to a geometric phase $\pi$. We have repeated this measurement for different durations $\plen$ of the middle coupling pulse. As we keep $\delta\varphi=2\pi$, this results in the same geometric loop being traced out at different speeds. For each measurement, we extract the phase $\gamma_0^-$ from the shift between the two Ramsey patterns and plot it versus $\plen$ [\fref{fig:vBp_Ramsey}(C), circles]. The data are clustered around the value $\pi$, confirming that $\gamma_0^-$ is largely independent of the rate at which we sweep $\varphi$ as long as the evolution stays adiabatic. This is a strong indication of the geometric character of $\gamma_0^-$. For the fastest pulses considered, we see systematic deviations from the value $\pi$. This behavior must be expected as the speed is increased, due to the breakdown of the adiabatic assumption. In the present case, the adiabaticity parameter can be written as $A=\pi/ (\coup \plen) = 0.52~\mu\rm{s}/\plen$. The crossover between adiabatic and nonadiabatic dynamics is expected when $A\approx1$ and $\plen\approx0.5~\mu\rm{s}$, in good agreement with the data of Fig.~3(C).

Using the same technique, we measure the phase acquired by the other eigenstate $\ket{\Psi_0^+}$. By adding a phase shift of $\pi$ to the coupling pulse, we turn $\ket{\Psi_0^-}$ into $\ket{\Psi_0^+}$, as the pseudospin is now aligned opposite to the effective magnetic field. The resulting geometric phase $\gamma_0^+$ [\fref{fig:vBp_Ramsey}(C), squares] follows a similar trend as $\gamma_0^-$, approaching $\pi$ in the adiabatic limit and deviating at shorter pulse durations. In addition, we consider the state $\ket{f0}$, for which the field mode is initially in the vacuum state. To prepare and measure $\ket{f0}$, we omit the first and third coupling pulses. As $\ket{f0}$ is not an eigenstate of \eqref{eq:Hquant}, we select only those pulse durations $\plen=\pi k/\coup$, with integer $k$, that give rise to a cyclic evolution. The resulting series [\fref{fig:vBp_Ramsey}(C), diamonds], in agreement with the other two, provides direct evidence of the vacuum-induced Berry phase.
Finally, we prepare an $n$-photon Fock state in the cavity and measure the phases acquired by the states $\eigens$ and $\ket{f,\nph}$ [\fref{fig:vBp_Ramsey}(D)].
The mean geometric phase, averaged over different states and different photon numbers $\nph=0,\ldots,4$, is $\mean{\gamma}=(3.1 \pm 0.2)\approx \pi$.
We thus conclude that, at resonance, the Berry phase is essentially independent of the photon number in the cavity.

\bigskip

\begin{figure}
\centering
	\includegraphics[width=\mywidth]
	{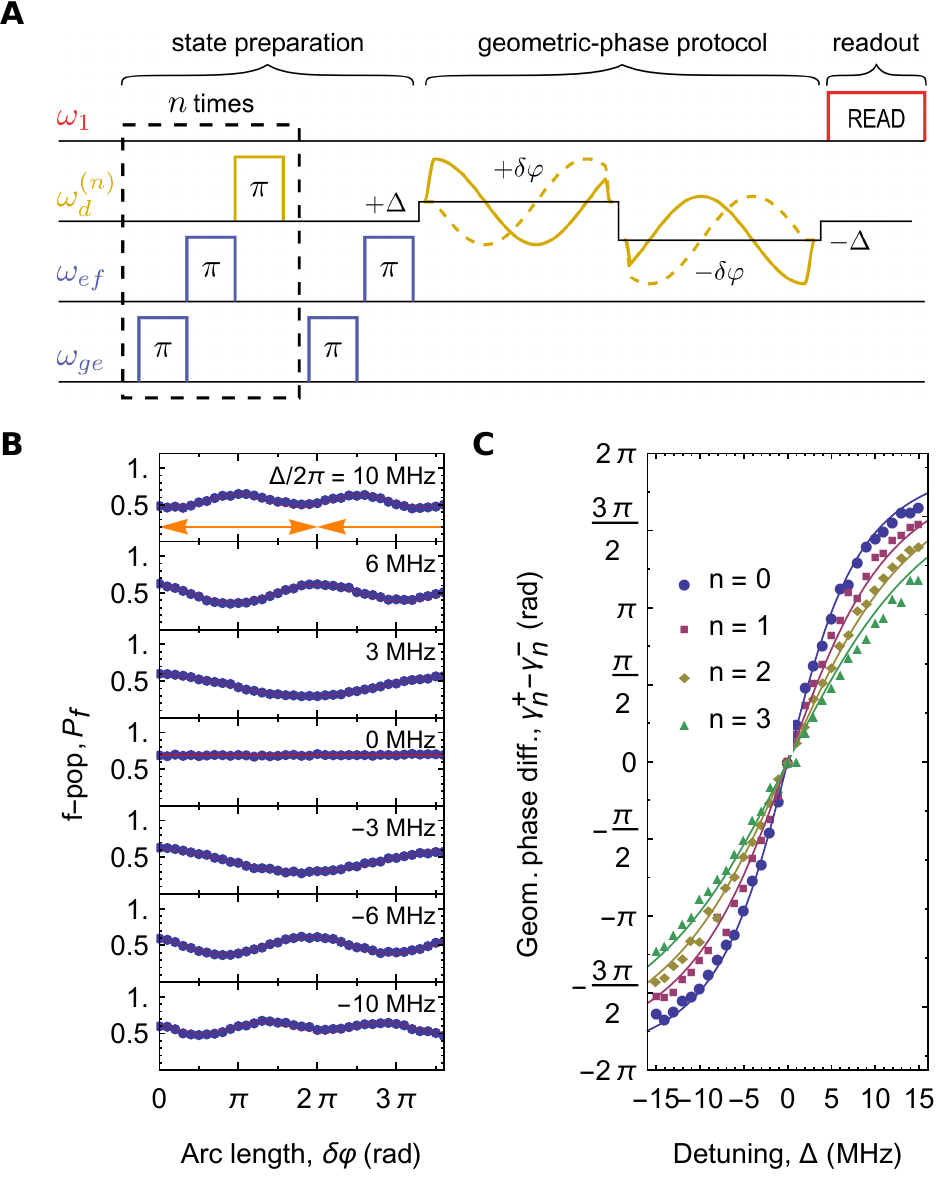}
	\caption{\textbf{Vacuum-induced Berry phase: finite detuning.}
\pA	Pulse sequence to detect the geometric phase difference accumulated between states $\ket{\Psi_\nph^-}$ and $\ket{\Psi_\nph^+}$ at finite detuning $\Delta$. We first prepare the state $\ket{f,\nph}$, which is a superposition of $\ket{\Psi_\nph^-}$ and $\ket{\Psi_\nph^+}$. Then we turn on the coupling and vary its phase by an amount $\delta\varphi$. We repeat this operation twice, the second time with an opposite detuning $-\Delta$, an opposite phase variation $-\delta\varphi$, and a $\pi$ phase shift. This sequence results in dynamic-phase cancellation, while the different geometric phases accumulated by $\ket{\Psi_\nph^\pm}$ can be detected as a population transfer away from the state $\ket{f,\nph}$.
\pB
Oscillations in the $f$-state population $P_f$ as a function of the phase displacement $\delta\varphi$, for selected values of the detuning $\Delta$ (circles: data, solid lines: sine fit) and $n=0$ photons in the cavity. The oscillation phase when $\delta\varphi=2\pi$ corresponds to the accumulated geometric phase in a single closed loop (whose extent is indicated by an orange line). \pC Geometric phase difference, $\gamma^+_n-\gamma_n^-$, versus detuning $\Delta$ for different photon numbers $\nph$ (symbols). The solid lines are a simultaneous fit of the model expression, Eq.~\eqref{eq:gammadtn}, to all data sets, with the coupling constant $g$ as the only fit parameter.
}
	\label{fig:vBp_detuned}
\end{figure}

%Results: detuned case
In contrast to the resonant case, a photon-number-dependent geometric phase is to be expected at finite detuning $\dtn$ between the atom and the field, as in that case the enclosed solid angle depends on the ratio $\dtn/(\coup\sqrt{n+1})$ [Fig.~1(D)].
Furthermore, according to Eq.~\eqref{eq:gammadtn}, the two eigenstates $\ket{\Psi_\nph^-}$ and $\ket{\Psi_\nph^+}$ acquire different phases: $\gamma_n^- \neq \gamma_n^+$ for $\dtn \neq 0$.
To measure the \textit{relative} geometric phase between $\ket{\Psi_\nph^-}$ and $\ket{\Psi_\nph^+}$ at arbitrary detuning, we use the pulse sequence described in Fig.~4(A).
First of all, we notice that for a generic $\Delta$ the state $\ket{f,\nph}=\alpha(\Delta) \ket{\Psi_\nph^-}+\beta(\Delta) \ket{\Psi_\nph^+}$ is a superposition of $\eigens$; as such, $\ket{f,\nph}$ can be directly used for Ramsey interferometry. The coefficients $\alpha(\Delta)$ and $\beta(\Delta)$ determine the visibility of the interference pattern.
As a measurement based on $\ket{f,\nph}$ only involves the two states $\eigens$, it allows us to use a spin--echo technique to cancel out the dynamic phase. While a spin-echo is typically implemented by applying an inverting $\pi$ pulse, here we prefer to engineer the effective Hamiltonian \eqref{eq:Hquant} so that the states $\eigens$ are effectively swapped during the second half of the evolution. This is accomplished by repeating the phase sweep with an opposite detuning,
an opposite phase variation,
and a phase shift of $\pi$ [see Fig.~4(A) and \cite{vBp_suppl}].
Finally, instead of varying the phase $\varphi$ by a full cycle ($\delta\varphi=2\pi$), we vary it by a fraction $\delta\varphi/2\pi$ of the full cycle. We repeat the measurement for incremental values of $\delta\varphi$ and record the corresponding $f$-state population $P_f$ at the end of the sequence. This protocol, based on a noncyclic geometric phase \cite{Samuel1988}, admits a similar geometric interpretation as in Fig.~1(D), provided the open ends of the paths described by $\eigens$ are connected to the initial state $\ket{f,n}$ by geodesic lines \cite{Samuel1988, vBp_suppl}. With this prescription, one finds that the acquired geometric phase is a linear function of $\delta\varphi$.

In \fref{fig:vBp_detuned}(B) we plot representative traces of $P_f$ versus $\delta\varphi$, for $\nph=0$ and different values of the detuning $\dtn$. The experimental data (dots) are fitted to sinusoidal oscillations (solid lines). The acquired geometric phase $\gamma$ after a full cycle ($\delta\varphi=2\pi$) is related to the frequency $f$ of the oscillations (with respect to $\delta\varphi$) by $\gamma= \pi f$. No oscillations are observed for $\dtn=0$. This is in good agreement with the results of \fref{fig:vBp_Ramsey}: at resonance, both states $\eigens$ acquire the same phase. As we move away from resonance, we observe oscillations of increasing frequency, indicating the accumulation of a geometric phase. The visibility of the oscillations decreases at higher detunings, due to our choice of $\ket{f0}$ as the reference state \cite{vBp_suppl}.
In \fref{fig:vBp_detuned}(C) we plot the geometric phase difference $(\gamma_\nph^+-\gamma_\nph^-)$ versus the detuning $\Delta$. Different symbols correspond to different photon numbers $\nph=0,1,2,3$.
We simultaneously fit our model expression, Eq.~\eqref{eq:gammadtn}, to all data sets (solid lines), with the coupling constant $g$ as the only fit parameter. The data are in good quantitative agreement with the model, with deviations of the order of a few percent at large detunings and higher photon numbers. From the global fit we extract the value $\coup/2\pi=(4.49\pm 0.03)~\rm{MHz}$. For comparison, an independent estimation based on Rabi oscillations gives $\coup/2\pi=(4.12\pm 0.06)~\rm{MHz}$ \cite{vBp_suppl}. We attribute the $8\%$ discrepancy between these two values to frequency-dependent attenuation in our input line (which includes a mixer and a room-temperature amplifier) as well as to higher-order transitions in our atom-cavity system, not accounted for in our model.

% Conclusions
The Berry phase induced by a quantized field can be thought of as a nontrivial combination of the geometric phase acquired by a quantum two-level system \cite{Leek2007} and that acquired by a harmonic oscillator \cite{Pechal2012}. Our experiments provide clear evidence of this phase, thus putting the theory predictions in Ref.~\cite{Fuentes-Guridi2002} on a solid empirical basis. The techniques demonstrated here may open new avenues for the geometric manipulation of atom-cavity systems, including geometric control of cavity states \cite{Vlastakis2013,Albert2015,Heeres2015} and cavity-assisted holonomic gates \cite{Abdumalikov2013}.
% ,Feng2013,Zu2014
For instance, our pulse scheme can be directly exploited to impart
a geometric phase onto specific Fock states in the cavity, similarly
to the results presented in a previous study \cite{Heeres2015}. In our case, two consecutive,
phase-shifted p pulses on the $\ket{f,n}$--$\ket{g,n+1}$ transition
realize a Fock state–selective phase gate in a time $\pi/\sqrt{n} g$, where
$g$ is the tunable coupling. Different Fock states can be simultaneously
addressed by exploiting the cavity-induced Stark shift on the $\ket{f,n}$--$\ket{g,n+1}$ transition, which is about 15 MHz in our system (see the
Supplementary Materials). As a further application, the tunable coupling
could be used to induce a cavity-mediated interaction between two
transmons, paving the way for the realization of a two-qubit geometric
gate based on non-Abelian holonomies \cite{Abdumalikov2013}.

\textbf{Acknowledgments} -- We thank Y. Salath\'e for technical assistance and J. Larson and E. Sjoqvist for helpful discussions. This work was supported by the Swiss National Science Foundation (SNF, project no.~150046).

%\bibliography{Biblio-vBp}   

\begin{thebibliography}{32}%
\makeatletter
\providecommand \@ifxundefined [1]{%
 \@ifx{#1\undefined}
}%
\providecommand \@ifnum [1]{%
 \ifnum #1\expandafter \@firstoftwo
 \else \expandafter \@secondoftwo
 \fi
}%
\providecommand \@ifx [1]{%
 \ifx #1\expandafter \@firstoftwo
 \else \expandafter \@secondoftwo
 \fi
}%
\providecommand \natexlab [1]{#1}%
\providecommand \enquote  [1]{``#1''}%
\providecommand \bibnamefont  [1]{#1}%
\providecommand \bibfnamefont [1]{#1}%
\providecommand \citenamefont [1]{#1}%
\providecommand \href@noop [0]{\@secondoftwo}%
\providecommand \href [0]{\begingroup \@sanitize@url \@href}%
\providecommand \@href[1]{\@@startlink{#1}\@@href}%
\providecommand \@@href[1]{\endgroup#1\@@endlink}%
\providecommand \@sanitize@url [0]{\catcode `\\12\catcode `\$12\catcode
  `\&12\catcode `\#12\catcode `\^12\catcode `\_12\catcode `\%12\relax}%
\providecommand \@@startlink[1]{}%
\providecommand \@@endlink[0]{}%
\providecommand \url  [0]{\begingroup\@sanitize@url \@url }%
\providecommand \@url [1]{\endgroup\@href {#1}{\urlprefix }}%
\providecommand \urlprefix  [0]{URL }%
\providecommand \Eprint [0]{\href }%
\providecommand \doibase [0]{http://dx.doi.org/}%
\providecommand \selectlanguage [0]{\@gobble}%
\providecommand \bibinfo  [0]{\@secondoftwo}%
\providecommand \bibfield  [0]{\@secondoftwo}%
\providecommand \translation [1]{[#1]}%
\providecommand \BibitemOpen [0]{}%
\providecommand \bibitemStop [0]{}%
\providecommand \bibitemNoStop [0]{.\EOS\space}%
\providecommand \EOS [0]{\spacefactor3000\relax}%
\providecommand \BibitemShut  [1]{\csname bibitem#1\endcsname}%
\let\auto@bib@innerbib\@empty
%</preamble>
\bibitem [{\citenamefont {Xiao}\ \emph {et~al.}(2010)\citenamefont {Xiao},
  \citenamefont {Chang},\ and\ \citenamefont {Niu}}]{Xiao2010}%
  \BibitemOpen
  \bibfield  {author} {\bibinfo {author} {\bibfnamefont {D.}~\bibnamefont
  {Xiao}}, \bibinfo {author} {\bibfnamefont {M.-C.}\ \bibnamefont {Chang}}, \
  and\ \bibinfo {author} {\bibfnamefont {Q.}~\bibnamefont {Niu}},\ }\href
  {\doibase 10.1103/RevModPhys.82.1959} {\bibfield  {journal} {\bibinfo
  {journal} {Reviews of Modern Physics}\ }\textbf {\bibinfo {volume} {82}},\
  \bibinfo {pages} {1959} (\bibinfo {year} {2010})}\BibitemShut {NoStop}%
\bibitem [{\citenamefont {Thouless}\ \emph {et~al.}(1982)\citenamefont
  {Thouless}, \citenamefont {Kohmoto}, \citenamefont {Nightingale},\ and\
  \citenamefont {den Nijs}}]{Thouless1982}%
  \BibitemOpen
  \bibfield  {author} {\bibinfo {author} {\bibfnamefont {D.~J.}\ \bibnamefont
  {Thouless}}, \bibinfo {author} {\bibfnamefont {M.}~\bibnamefont {Kohmoto}},
  \bibinfo {author} {\bibfnamefont {M.~P.}\ \bibnamefont {Nightingale}}, \ and\
  \bibinfo {author} {\bibfnamefont {M.}~\bibnamefont {den Nijs}},\ }\href
  {http://journals.aps.org/prl/abstract/10.1103/PhysRevLett.49.405} {\bibfield
  {journal} {\bibinfo  {journal} {Physical Review Letters}\ }\textbf {\bibinfo
  {volume} {49}},\ \bibinfo {pages} {405} (\bibinfo {year} {1982})}\BibitemShut
  {NoStop}%
\bibitem [{\citenamefont {Hasan}\ and\ \citenamefont {Kane}(2010)}]{Hasan2010}%
  \BibitemOpen
  \bibfield  {author} {\bibinfo {author} {\bibfnamefont {M.~Z.}\ \bibnamefont
  {Hasan}}\ and\ \bibinfo {author} {\bibfnamefont {C.~L.}\ \bibnamefont
  {Kane}},\ }\href {\doibase 10.1103/RevModPhys.82.3045} {\bibfield  {journal}
  {\bibinfo  {journal} {Reviews of Modern Physics}\ }\textbf {\bibinfo {volume}
  {82}},\ \bibinfo {pages} {3045} (\bibinfo {year} {2010})}\BibitemShut
  {NoStop}%
\bibitem [{\citenamefont {Qi}\ and\ \citenamefont {Zhang}(2011)}]{Qi2011}%
  \BibitemOpen
  \bibfield  {author} {\bibinfo {author} {\bibfnamefont {X.~L.}\ \bibnamefont
  {Qi}}\ and\ \bibinfo {author} {\bibfnamefont {S.~C.}\ \bibnamefont {Zhang}},\
  }\href {\doibase 10.1103/RevModPhys.83.1057} {\bibfield  {journal} {\bibinfo
  {journal} {Reviews of Modern Physics}\ }\textbf {\bibinfo {volume} {83}},\
  \bibinfo {pages} {1057} (\bibinfo {year} {2011})}\BibitemShut {NoStop}%
\bibitem [{\citenamefont {Zanardi}\ and\ \citenamefont
  {Rasetti}(1999)}]{Zanardi1999}%
  \BibitemOpen
  \bibfield  {author} {\bibinfo {author} {\bibfnamefont {P.}~\bibnamefont
  {Zanardi}}\ and\ \bibinfo {author} {\bibfnamefont {M.}~\bibnamefont
  {Rasetti}},\ }\href {\doibase 10.1016/S0375-9601(99)00803-8} {\bibfield
  {journal} {\bibinfo  {journal} {Physics Letters A}\ }\textbf {\bibinfo
  {volume} {264}},\ \bibinfo {pages} {94} (\bibinfo {year} {1999})}\BibitemShut
  {NoStop}%
\bibitem [{\citenamefont {Sj\"{o}qvist}(2008)}]{Sjoqvist2008}%
  \BibitemOpen
  \bibfield  {author} {\bibinfo {author} {\bibfnamefont {E.}~\bibnamefont
  {Sj\"{o}qvist}},\ }\href {http://link.aps.org/doi/10.1103/Physics.1.35}
  {\bibfield  {journal} {\bibinfo  {journal} {Physics}\ }\textbf {\bibinfo
  {volume} {1}} (\bibinfo {year} {2008})}\BibitemShut {NoStop}%
\bibitem [{\citenamefont {Berry}(1984)}]{Berry1984}%
  \BibitemOpen
  \bibfield  {author} {\bibinfo {author} {\bibfnamefont {M.~V.}\ \bibnamefont
  {Berry}},\ }\href@noop {} {\bibfield  {journal} {\bibinfo  {journal}
  {Proceedings of the Royal Society of London A}\ }\textbf {\bibinfo {volume}
  {392}},\ \bibinfo {pages} {45} (\bibinfo {year} {1984})}\BibitemShut
  {NoStop}%
\bibitem [{\citenamefont {Fuentes-Guridi}\ \emph {et~al.}(2002)\citenamefont
  {Fuentes-Guridi}, \citenamefont {Carollo}, \citenamefont {Bose},\ and\
  \citenamefont {Vedral}}]{Fuentes-Guridi2002}%
  \BibitemOpen
  \bibfield  {author} {\bibinfo {author} {\bibfnamefont {I.}~\bibnamefont
  {Fuentes-Guridi}}, \bibinfo {author} {\bibfnamefont {A.}~\bibnamefont
  {Carollo}}, \bibinfo {author} {\bibfnamefont {S.}~\bibnamefont {Bose}}, \
  and\ \bibinfo {author} {\bibfnamefont {V.}~\bibnamefont {Vedral}},\ }\href
  {\doibase 10.1103/PhysRevLett.89.220404} {\bibfield  {journal} {\bibinfo
  {journal} {Physical Review Letters}\ }\textbf {\bibinfo {volume} {89}},\
  \bibinfo {pages} {220404} (\bibinfo {year} {2002})}\BibitemShut {NoStop}%
\bibitem [{\citenamefont {Liu}\ \emph {et~al.}(2011)\citenamefont {Liu},
  \citenamefont {Feng},\ and\ \citenamefont {Wang}}]{Liu2011a}%
  \BibitemOpen
  \bibfield  {author} {\bibinfo {author} {\bibfnamefont {T.}~\bibnamefont
  {Liu}}, \bibinfo {author} {\bibfnamefont {M.}~\bibnamefont {Feng}}, \ and\
  \bibinfo {author} {\bibfnamefont {K.}~\bibnamefont {Wang}},\ }\href {\doibase
  10.1103/PhysRevA.84.062109} {\bibfield  {journal} {\bibinfo  {journal}
  {Physical Review A}\ }\textbf {\bibinfo {volume} {84}},\ \bibinfo {pages}
  {062109} (\bibinfo {year} {2011})}\BibitemShut {NoStop}%
\bibitem [{\citenamefont {Larson}(2012)}]{Larson2012}%
  \BibitemOpen
  \bibfield  {author} {\bibinfo {author} {\bibfnamefont {J.}~\bibnamefont
  {Larson}},\ }\href {\doibase 10.1103/PhysRevLett.108.033601} {\bibfield
  {journal} {\bibinfo  {journal} {Physical Review Letters}\ }\textbf {\bibinfo
  {volume} {108}},\ \bibinfo {pages} {033601} (\bibinfo {year}
  {2012})}\BibitemShut {NoStop}%
\bibitem [{\citenamefont {Wang}\ \emph {et~al.}(2015)\citenamefont {Wang},
  \citenamefont {Wei},\ and\ \citenamefont {Liang}}]{Wang2015}%
  \BibitemOpen
  \bibfield  {author} {\bibinfo {author} {\bibfnamefont {M.}~\bibnamefont
  {Wang}}, \bibinfo {author} {\bibfnamefont {L.}~\bibnamefont {Wei}}, \ and\
  \bibinfo {author} {\bibfnamefont {J.}~\bibnamefont {Liang}},\ }\href
  {\doibase 10.1016/j.physleta.2015.02.006} {\bibfield  {journal} {\bibinfo
  {journal} {Physics Letters A}\ }\textbf {\bibinfo {volume} {379}},\ \bibinfo
  {pages} {1087} (\bibinfo {year} {2015})}\BibitemShut {NoStop}%
\bibitem [{\citenamefont {Falci}\ \emph {et~al.}(2000)\citenamefont {Falci},
  \citenamefont {Fazio}, \citenamefont {Palma}, \citenamefont {Siewert},\ and\
  \citenamefont {Vedral}}]{Falci2000}%
  \BibitemOpen
  \bibfield  {author} {\bibinfo {author} {\bibfnamefont {G.}~\bibnamefont
  {Falci}}, \bibinfo {author} {\bibfnamefont {R.}~\bibnamefont {Fazio}},
  \bibinfo {author} {\bibfnamefont {G.~M.}\ \bibnamefont {Palma}}, \bibinfo
  {author} {\bibfnamefont {J.}~\bibnamefont {Siewert}}, \ and\ \bibinfo
  {author} {\bibfnamefont {V.}~\bibnamefont {Vedral}},\ }\href {\doibase
  10.1038/35030052} {\bibfield  {journal} {\bibinfo  {journal} {Nature}\
  }\textbf {\bibinfo {volume} {407}},\ \bibinfo {pages} {355} (\bibinfo {year}
  {2000})}\BibitemShut {NoStop}%
\bibitem [{\citenamefont {Leek}\ \emph {et~al.}(2007)\citenamefont {Leek},
  \citenamefont {Fink}, \citenamefont {Blais}, \citenamefont {Bianchetti},
  \citenamefont {G\"{o}ppl}, \citenamefont {Gambetta}, \citenamefont
  {Schuster}, \citenamefont {Frunzio}, \citenamefont {Schoelkopf},\ and\
  \citenamefont {Wallraff}}]{Leek2007}%
  \BibitemOpen
  \bibfield  {author} {\bibinfo {author} {\bibfnamefont {P.~J.}\ \bibnamefont
  {Leek}}, \bibinfo {author} {\bibfnamefont {J.~M.}\ \bibnamefont {Fink}},
  \bibinfo {author} {\bibfnamefont {A.}~\bibnamefont {Blais}}, \bibinfo
  {author} {\bibfnamefont {R.}~\bibnamefont {Bianchetti}}, \bibinfo {author}
  {\bibfnamefont {M.}~\bibnamefont {G\"{o}ppl}}, \bibinfo {author}
  {\bibfnamefont {J.~M.}\ \bibnamefont {Gambetta}}, \bibinfo {author}
  {\bibfnamefont {D.~I.}\ \bibnamefont {Schuster}}, \bibinfo {author}
  {\bibfnamefont {L.}~\bibnamefont {Frunzio}}, \bibinfo {author} {\bibfnamefont
  {R.~J.}\ \bibnamefont {Schoelkopf}}, \ and\ \bibinfo {author} {\bibfnamefont
  {A.}~\bibnamefont {Wallraff}},\ }\href {\doibase 10.1126/science.1149858}
  {\bibfield  {journal} {\bibinfo  {journal} {Science}\ }\textbf {\bibinfo
  {volume} {318}},\ \bibinfo {pages} {1889} (\bibinfo {year}
  {2007})}\BibitemShut {NoStop}%
\bibitem [{\citenamefont {M\"{o}tt\"{o}nen}\ \emph {et~al.}(2008)\citenamefont
  {M\"{o}tt\"{o}nen}, \citenamefont {Vartiainen},\ and\ \citenamefont
  {Pekola}}]{Mottonen2008}%
  \BibitemOpen
  \bibfield  {author} {\bibinfo {author} {\bibfnamefont {M.}~\bibnamefont
  {M\"{o}tt\"{o}nen}}, \bibinfo {author} {\bibfnamefont {J.~J.}\ \bibnamefont
  {Vartiainen}}, \ and\ \bibinfo {author} {\bibfnamefont {J.~P.}\ \bibnamefont
  {Pekola}},\ }\href {\doibase 10.1103/PhysRevLett.100.177201} {\bibfield
  {journal} {\bibinfo  {journal} {Physical Review Letters}\ }\textbf {\bibinfo
  {volume} {100}},\ \bibinfo {pages} {177201} (\bibinfo {year}
  {2008})}\BibitemShut {NoStop}%
\bibitem [{\citenamefont {Pechal}\ \emph {et~al.}(2012)\citenamefont {Pechal},
  \citenamefont {Berger}, \citenamefont {Abdumalikov}, \citenamefont {Fink},
  \citenamefont {Mlynek}, \citenamefont {Steffen}, \citenamefont {Wallraff},\
  and\ \citenamefont {Filipp}}]{Pechal2012}%
  \BibitemOpen
  \bibfield  {author} {\bibinfo {author} {\bibfnamefont {M.}~\bibnamefont
  {Pechal}}, \bibinfo {author} {\bibfnamefont {S.}~\bibnamefont {Berger}},
  \bibinfo {author} {\bibfnamefont {A.~A.}\ \bibnamefont {Abdumalikov}},
  \bibinfo {author} {\bibfnamefont {J.~M.}\ \bibnamefont {Fink}}, \bibinfo
  {author} {\bibfnamefont {J.~a.}\ \bibnamefont {Mlynek}}, \bibinfo {author}
  {\bibfnamefont {L.}~\bibnamefont {Steffen}}, \bibinfo {author} {\bibfnamefont
  {A.}~\bibnamefont {Wallraff}}, \ and\ \bibinfo {author} {\bibfnamefont
  {S.}~\bibnamefont {Filipp}},\ }\href {\doibase
  10.1103/PhysRevLett.108.170401} {\bibfield  {journal} {\bibinfo  {journal}
  {Physical Review Letters}\ }\textbf {\bibinfo {volume} {108}},\ \bibinfo
  {pages} {170401} (\bibinfo {year} {2012})}\BibitemShut {NoStop}%
\bibitem [{\citenamefont {Abdumalikov}\ \emph {et~al.}(2013)\citenamefont
  {Abdumalikov}, \citenamefont {Fink}, \citenamefont {Juliusson}, \citenamefont
  {Pechal}, \citenamefont {Berger}, \citenamefont {Wallraff},\ and\
  \citenamefont {Filipp}}]{Abdumalikov2013}%
  \BibitemOpen
  \bibfield  {author} {\bibinfo {author} {\bibfnamefont {A.~A.}\ \bibnamefont
  {Abdumalikov}}, \bibinfo {author} {\bibfnamefont {J.~M.}\ \bibnamefont
  {Fink}}, \bibinfo {author} {\bibfnamefont {K.}~\bibnamefont {Juliusson}},
  \bibinfo {author} {\bibfnamefont {M.}~\bibnamefont {Pechal}}, \bibinfo
  {author} {\bibfnamefont {S.}~\bibnamefont {Berger}}, \bibinfo {author}
  {\bibfnamefont {A.}~\bibnamefont {Wallraff}}, \ and\ \bibinfo {author}
  {\bibfnamefont {S.}~\bibnamefont {Filipp}},\ }\href {\doibase
  10.1038/nature12010} {\bibfield  {journal} {\bibinfo  {journal} {Nature}\
  }\textbf {\bibinfo {volume} {496}},\ \bibinfo {pages} {482} (\bibinfo {year}
  {2013})}\BibitemShut {NoStop}%
\bibitem [{\citenamefont {Berger}\ \emph {et~al.}(2013)\citenamefont {Berger},
  \citenamefont {Pechal}, \citenamefont {Abdumalikov}, \citenamefont {Eichler},
  \citenamefont {Steffen}, \citenamefont {Fedorov}, \citenamefont {Wallraff},\
  and\ \citenamefont {Filipp}}]{Berger2013}%
  \BibitemOpen
  \bibfield  {author} {\bibinfo {author} {\bibfnamefont {S.}~\bibnamefont
  {Berger}}, \bibinfo {author} {\bibfnamefont {M.}~\bibnamefont {Pechal}},
  \bibinfo {author} {\bibfnamefont {A.~A.}\ \bibnamefont {Abdumalikov}},
  \bibinfo {author} {\bibfnamefont {C.}~\bibnamefont {Eichler}}, \bibinfo
  {author} {\bibfnamefont {L.}~\bibnamefont {Steffen}}, \bibinfo {author}
  {\bibfnamefont {A.}~\bibnamefont {Fedorov}}, \bibinfo {author} {\bibfnamefont
  {A.}~\bibnamefont {Wallraff}}, \ and\ \bibinfo {author} {\bibfnamefont
  {S.}~\bibnamefont {Filipp}},\ }\href {\doibase 10.1103/PhysRevA.87.060303}
  {\bibfield  {journal} {\bibinfo  {journal} {Physical Review A}\ }\textbf
  {\bibinfo {volume} {87}},\ \bibinfo {pages} {060303} (\bibinfo {year}
  {2013})}\BibitemShut {NoStop}%
\bibitem [{\citenamefont {Roushan}\ \emph {et~al.}(2014)\citenamefont
  {Roushan}, \citenamefont {Neill}, \citenamefont {Chen}, \citenamefont
  {Kolodrubetz}, \citenamefont {Quintana}, \citenamefont {Leung}, \citenamefont
  {Fang}, \citenamefont {Barends}, \citenamefont {Campbell}, \citenamefont
  {Chen}, \citenamefont {Chiaro}, \citenamefont {Dunsworth}, \citenamefont
  {Jeffrey}, \citenamefont {Kelly}, \citenamefont {Megrant}, \citenamefont
  {Mutus}, \citenamefont {{O' Malley}}, \citenamefont {Sank}, \citenamefont
  {Vainsencher}, \citenamefont {Wenner}, \citenamefont {White}, \citenamefont
  {Polkovnikov}, \citenamefont {Cleland},\ and\ \citenamefont
  {Martinis}}]{Roushan2014}%
  \BibitemOpen
  \bibfield  {author} {\bibinfo {author} {\bibfnamefont {P.}~\bibnamefont
  {Roushan}}, \bibinfo {author} {\bibfnamefont {C.}~\bibnamefont {Neill}},
  \bibinfo {author} {\bibfnamefont {Y.}~\bibnamefont {Chen}}, \bibinfo {author}
  {\bibfnamefont {M.}~\bibnamefont {Kolodrubetz}}, \bibinfo {author}
  {\bibfnamefont {C.}~\bibnamefont {Quintana}}, \bibinfo {author}
  {\bibfnamefont {N.}~\bibnamefont {Leung}}, \bibinfo {author} {\bibfnamefont
  {M.}~\bibnamefont {Fang}}, \bibinfo {author} {\bibfnamefont {R.}~\bibnamefont
  {Barends}}, \bibinfo {author} {\bibfnamefont {B.}~\bibnamefont {Campbell}},
  \bibinfo {author} {\bibfnamefont {Z.}~\bibnamefont {Chen}}, \bibinfo {author}
  {\bibfnamefont {B.}~\bibnamefont {Chiaro}}, \bibinfo {author} {\bibfnamefont
  {A.}~\bibnamefont {Dunsworth}}, \bibinfo {author} {\bibfnamefont
  {E.}~\bibnamefont {Jeffrey}}, \bibinfo {author} {\bibfnamefont
  {J.}~\bibnamefont {Kelly}}, \bibinfo {author} {\bibfnamefont
  {A.}~\bibnamefont {Megrant}}, \bibinfo {author} {\bibfnamefont
  {J.}~\bibnamefont {Mutus}}, \bibinfo {author} {\bibfnamefont {P.~J.~J.}\
  \bibnamefont {{O' Malley}}}, \bibinfo {author} {\bibfnamefont
  {D.}~\bibnamefont {Sank}}, \bibinfo {author} {\bibfnamefont {A.}~\bibnamefont
  {Vainsencher}}, \bibinfo {author} {\bibfnamefont {J.}~\bibnamefont {Wenner}},
  \bibinfo {author} {\bibfnamefont {T.}~\bibnamefont {White}}, \bibinfo
  {author} {\bibfnamefont {A.}~\bibnamefont {Polkovnikov}}, \bibinfo {author}
  {\bibfnamefont {A.~N.}\ \bibnamefont {Cleland}}, \ and\ \bibinfo {author}
  {\bibfnamefont {J.~M.}\ \bibnamefont {Martinis}},\ }\href {\doibase
  10.1038/nature13891} {\bibfield  {journal} {\bibinfo  {journal} {Nature}\
  }\textbf {\bibinfo {volume} {515}},\ \bibinfo {pages} {241} (\bibinfo {year}
  {2014})}\BibitemShut {NoStop}%
\bibitem [{\citenamefont {Schroer}\ \emph {et~al.}(2014)\citenamefont
  {Schroer}, \citenamefont {Kolodrubetz}, \citenamefont {Kindel}, \citenamefont
  {Sandberg}, \citenamefont {Gao}, \citenamefont {Vissers}, \citenamefont
  {Pappas}, \citenamefont {Polkovnikov},\ and\ \citenamefont
  {Lehnert}}]{Schroer2014}%
  \BibitemOpen
  \bibfield  {author} {\bibinfo {author} {\bibfnamefont {M.~D.}\ \bibnamefont
  {Schroer}}, \bibinfo {author} {\bibfnamefont {M.~H.}\ \bibnamefont
  {Kolodrubetz}}, \bibinfo {author} {\bibfnamefont {W.~F.}\ \bibnamefont
  {Kindel}}, \bibinfo {author} {\bibfnamefont {M.}~\bibnamefont {Sandberg}},
  \bibinfo {author} {\bibfnamefont {J.}~\bibnamefont {Gao}}, \bibinfo {author}
  {\bibfnamefont {M.~R.}\ \bibnamefont {Vissers}}, \bibinfo {author}
  {\bibfnamefont {D.~P.}\ \bibnamefont {Pappas}}, \bibinfo {author}
  {\bibfnamefont {A.}~\bibnamefont {Polkovnikov}}, \ and\ \bibinfo {author}
  {\bibfnamefont {K.~W.}\ \bibnamefont {Lehnert}},\ }\href {\doibase
  10.1103/PhysRevLett.113.050402} {\bibfield  {journal} {\bibinfo  {journal}
  {Physical Review Letters}\ }\textbf {\bibinfo {volume} {113}},\ \bibinfo
  {pages} {050402} (\bibinfo {year} {2014})}\BibitemShut {NoStop}%
\bibitem [{\citenamefont {Jones}\ \emph {et~al.}(2000)\citenamefont {Jones},
  \citenamefont {Vedral}, \citenamefont {Ekert},\ and\ \citenamefont
  {Castagnoli}}]{Jones2000}%
  \BibitemOpen
  \bibfield  {author} {\bibinfo {author} {\bibfnamefont {J.~A.}\ \bibnamefont
  {Jones}}, \bibinfo {author} {\bibfnamefont {V.}~\bibnamefont {Vedral}},
  \bibinfo {author} {\bibfnamefont {A.}~\bibnamefont {Ekert}}, \ and\ \bibinfo
  {author} {\bibfnamefont {G.}~\bibnamefont {Castagnoli}},\ }\href@noop {}
  {\bibfield  {journal} {\bibinfo  {journal} {Nature}\ }\textbf {\bibinfo
  {volume} {403}},\ \bibinfo {pages} {869} (\bibinfo {year}
  {2000})}\BibitemShut {NoStop}%
\bibitem [{\citenamefont {Bose}\ \emph {et~al.}(2003)\citenamefont {Bose},
  \citenamefont {Carollo}, \citenamefont {Fuentes-Guridi}, \citenamefont
  {{Franca Santos}},\ and\ \citenamefont {Vedral}}]{Bose2003}%
  \BibitemOpen
  \bibfield  {author} {\bibinfo {author} {\bibfnamefont {S.}~\bibnamefont
  {Bose}}, \bibinfo {author} {\bibfnamefont {A.}~\bibnamefont {Carollo}},
  \bibinfo {author} {\bibfnamefont {I.}~\bibnamefont {Fuentes-Guridi}},
  \bibinfo {author} {\bibfnamefont {M.}~\bibnamefont {{Franca Santos}}}, \ and\
  \bibinfo {author} {\bibfnamefont {V.}~\bibnamefont {Vedral}},\ }\href
  {\doibase 10.1080/0950034031000069505} {\bibfield  {journal} {\bibinfo
  {journal} {Journal of Modern Optics}\ }\textbf {\bibinfo {volume} {50}},\
  \bibinfo {pages} {1175} (\bibinfo {year} {2003})}\BibitemShut {NoStop}%
\bibitem [{\citenamefont {Liu}\ \emph {et~al.}(2010)\citenamefont {Liu},
  \citenamefont {Wei}, \citenamefont {Jia},\ and\ \citenamefont
  {Liang}}]{Liu2010}%
  \BibitemOpen
  \bibfield  {author} {\bibinfo {author} {\bibfnamefont {Y.}~\bibnamefont
  {Liu}}, \bibinfo {author} {\bibfnamefont {L.~F.}\ \bibnamefont {Wei}},
  \bibinfo {author} {\bibfnamefont {W.~Z.}\ \bibnamefont {Jia}}, \ and\
  \bibinfo {author} {\bibfnamefont {J.~Q.}\ \bibnamefont {Liang}},\ }\href
  {\doibase 10.1103/PhysRevA.82.045801} {\bibfield  {journal} {\bibinfo
  {journal} {Physical Review A}\ }\textbf {\bibinfo {volume} {82}},\ \bibinfo
  {pages} {045801} (\bibinfo {year} {2010})}\BibitemShut {NoStop}%
\bibitem [{\citenamefont {Pechal}\ \emph {et~al.}(2014)\citenamefont {Pechal},
  \citenamefont {Huthmacher}, \citenamefont {Eichler}, \citenamefont
  {Zeytino\v{g}lu}, \citenamefont {Abdumalikov}, \citenamefont {Berger},
  \citenamefont {Wallraff},\ and\ \citenamefont {Filipp}}]{Pechal2014}%
  \BibitemOpen
  \bibfield  {author} {\bibinfo {author} {\bibfnamefont {M.}~\bibnamefont
  {Pechal}}, \bibinfo {author} {\bibfnamefont {L.}~\bibnamefont {Huthmacher}},
  \bibinfo {author} {\bibfnamefont {C.}~\bibnamefont {Eichler}}, \bibinfo
  {author} {\bibfnamefont {S.}~\bibnamefont {Zeytino\v{g}lu}}, \bibinfo
  {author} {\bibfnamefont {A.~A.}\ \bibnamefont {Abdumalikov}}, \bibinfo
  {author} {\bibfnamefont {S.}~\bibnamefont {Berger}}, \bibinfo {author}
  {\bibfnamefont {A.}~\bibnamefont {Wallraff}}, \ and\ \bibinfo {author}
  {\bibfnamefont {S.}~\bibnamefont {Filipp}},\ }\href {\doibase
  10.1103/PhysRevX.4.041010} {\bibfield  {journal} {\bibinfo  {journal}
  {Physical Review X}\ }\textbf {\bibinfo {volume} {4}},\ \bibinfo {pages}
  {041010} (\bibinfo {year} {2014})}\BibitemShut {NoStop}%
\bibitem [{\citenamefont {Zeytino\v{g}lu}\ \emph {et~al.}(2015)\citenamefont
  {Zeytino\v{g}lu}, \citenamefont {Pechal}, \citenamefont {Berger},
  \citenamefont {Abdumalikov}, \citenamefont {Wallraff},\ and\ \citenamefont
  {Filipp}}]{Zeytinoglu2015}%
  \BibitemOpen
  \bibfield  {author} {\bibinfo {author} {\bibfnamefont {S.}~\bibnamefont
  {Zeytino\v{g}lu}}, \bibinfo {author} {\bibfnamefont {M.}~\bibnamefont
  {Pechal}}, \bibinfo {author} {\bibfnamefont {S.}~\bibnamefont {Berger}},
  \bibinfo {author} {\bibfnamefont {A.~A.}\ \bibnamefont {Abdumalikov}},
  \bibinfo {author} {\bibfnamefont {A.}~\bibnamefont {Wallraff}}, \ and\
  \bibinfo {author} {\bibfnamefont {S.}~\bibnamefont {Filipp}},\ }\href
  {\doibase 10.1103/PhysRevA.91.043846} {\bibfield  {journal} {\bibinfo
  {journal} {Physical Review A}\ }\textbf {\bibinfo {volume} {91}},\ \bibinfo
  {pages} {43846} (\bibinfo {year} {2015})}\BibitemShut {NoStop}%
\bibitem [{\citenamefont {Paik}\ \emph {et~al.}(2011)\citenamefont {Paik},
  \citenamefont {Schuster}, \citenamefont {Bishop}, \citenamefont {Kirchmair},
  \citenamefont {Catelani}, \citenamefont {Sears}, \citenamefont {Johnson},
  \citenamefont {Reagor}, \citenamefont {Frunzio}, \citenamefont {Glazman},
  \citenamefont {Girvin}, \citenamefont {Devoret},\ and\ \citenamefont
  {Schoelkopf}}]{Paik2011}%
  \BibitemOpen
  \bibfield  {author} {\bibinfo {author} {\bibfnamefont {H.}~\bibnamefont
  {Paik}}, \bibinfo {author} {\bibfnamefont {D.~I.}\ \bibnamefont {Schuster}},
  \bibinfo {author} {\bibfnamefont {L.~S.}\ \bibnamefont {Bishop}}, \bibinfo
  {author} {\bibfnamefont {G.}~\bibnamefont {Kirchmair}}, \bibinfo {author}
  {\bibfnamefont {G.}~\bibnamefont {Catelani}}, \bibinfo {author}
  {\bibfnamefont {a.~P.}\ \bibnamefont {Sears}}, \bibinfo {author}
  {\bibfnamefont {B.~R.}\ \bibnamefont {Johnson}}, \bibinfo {author}
  {\bibfnamefont {M.~J.}\ \bibnamefont {Reagor}}, \bibinfo {author}
  {\bibfnamefont {L.}~\bibnamefont {Frunzio}}, \bibinfo {author} {\bibfnamefont
  {L.~I.}\ \bibnamefont {Glazman}}, \bibinfo {author} {\bibfnamefont {S.~M.}\
  \bibnamefont {Girvin}}, \bibinfo {author} {\bibfnamefont {M.}~\bibnamefont
  {Devoret}}, \ and\ \bibinfo {author} {\bibfnamefont {R.~J.}\ \bibnamefont
  {Schoelkopf}},\ }\href {\doibase 10.1103/PhysRevLett.107.240501} {\bibfield
  {journal} {\bibinfo  {journal} {Physical Review Letters}\ }\textbf {\bibinfo
  {volume} {107}},\ \bibinfo {pages} {240501} (\bibinfo {year}
  {2011})}\BibitemShut {NoStop}%
\bibitem [{\citenamefont {Bianchetti}\ \emph {et~al.}(2010)\citenamefont
  {Bianchetti}, \citenamefont {Filipp}, \citenamefont {Baur}, \citenamefont
  {Fink}, \citenamefont {Lang}, \citenamefont {Steffen}, \citenamefont
  {Boissonneault}, \citenamefont {Blais},\ and\ \citenamefont
  {Wallraff}}]{Bianchetti2010}%
  \BibitemOpen
  \bibfield  {author} {\bibinfo {author} {\bibfnamefont {R.}~\bibnamefont
  {Bianchetti}}, \bibinfo {author} {\bibfnamefont {S.}~\bibnamefont {Filipp}},
  \bibinfo {author} {\bibfnamefont {M.}~\bibnamefont {Baur}}, \bibinfo {author}
  {\bibfnamefont {J.~M.}\ \bibnamefont {Fink}}, \bibinfo {author}
  {\bibfnamefont {C.}~\bibnamefont {Lang}}, \bibinfo {author} {\bibfnamefont
  {L.}~\bibnamefont {Steffen}}, \bibinfo {author} {\bibfnamefont
  {M.}~\bibnamefont {Boissonneault}}, \bibinfo {author} {\bibfnamefont
  {A.}~\bibnamefont {Blais}}, \ and\ \bibinfo {author} {\bibfnamefont
  {A.}~\bibnamefont {Wallraff}},\ }\href {\doibase
  10.1103/PhysRevLett.105.223601} {\bibfield  {journal} {\bibinfo  {journal}
  {Physical Review Letters}\ }\textbf {\bibinfo {volume} {105}},\ \bibinfo
  {pages} {223601} (\bibinfo {year} {2010})}\BibitemShut {NoStop}%
\bibitem [{\citenamefont {Schuster}\ \emph {et~al.}(2007)\citenamefont
  {Schuster}, \citenamefont {Houck}, \citenamefont {Schreier}, \citenamefont
  {Wallraff}, \citenamefont {Gambetta}, \citenamefont {Blais}, \citenamefont
  {Frunzio}, \citenamefont {Majer}, \citenamefont {Johnson}, \citenamefont
  {Devoret}, \citenamefont {Girvin},\ and\ \citenamefont
  {Schoelkopf}}]{Schuster2007}%
  \BibitemOpen
\bibfield  {journal} {  }\bibfield  {author} {\bibinfo {author} {\bibfnamefont
  {D.~I.}\ \bibnamefont {Schuster}}, \bibinfo {author} {\bibfnamefont {A.~A.}\
  \bibnamefont {Houck}}, \bibinfo {author} {\bibfnamefont {J.~A.}\ \bibnamefont
  {Schreier}}, \bibinfo {author} {\bibfnamefont {A.}~\bibnamefont {Wallraff}},
  \bibinfo {author} {\bibfnamefont {J.~M.}\ \bibnamefont {Gambetta}}, \bibinfo
  {author} {\bibfnamefont {A.}~\bibnamefont {Blais}}, \bibinfo {author}
  {\bibfnamefont {L.}~\bibnamefont {Frunzio}}, \bibinfo {author} {\bibfnamefont
  {J.}~\bibnamefont {Majer}}, \bibinfo {author} {\bibfnamefont
  {B.}~\bibnamefont {Johnson}}, \bibinfo {author} {\bibfnamefont {M.~H.}\
  \bibnamefont {Devoret}}, \bibinfo {author} {\bibfnamefont {S.~M.}\
  \bibnamefont {Girvin}}, \ and\ \bibinfo {author} {\bibfnamefont {R.~J.}\
  \bibnamefont {Schoelkopf}},\ }\href {\doibase 10.1038/nature05461} {\bibfield
   {journal} {\bibinfo  {journal} {Nature}\ }\textbf {\bibinfo {volume}
  {445}},\ \bibinfo {pages} {515} (\bibinfo {year} {2007})}\BibitemShut
  {NoStop}%
\bibitem [{\citenamefont {Samuel}\ and\ \citenamefont
  {Bhandari}(1988)}]{Samuel1988}%
  \BibitemOpen
  \bibfield  {author} {\bibinfo {author} {\bibfnamefont {J.}~\bibnamefont
  {Samuel}}\ and\ \bibinfo {author} {\bibfnamefont {R.}~\bibnamefont
  {Bhandari}},\ }\href@noop {} {\bibfield  {journal} {\bibinfo  {journal}
  {Physical Review Letters}\ }\textbf {\bibinfo {volume} {60}},\ \bibinfo
  {pages} {2339} (\bibinfo {year} {1988})}\BibitemShut {NoStop}%
\bibitem [{\citenamefont {Vlastakis}\ \emph {et~al.}(2013)\citenamefont
  {Vlastakis}, \citenamefont {Kirchmair}, \citenamefont {Leghtas},
  \citenamefont {Nigg}, \citenamefont {Frunzio}, \citenamefont {Girvin},
  \citenamefont {Mirrahimi}, \citenamefont {Devoret},\ and\ \citenamefont
  {Schoelkopf}}]{Vlastakis2013}%
  \BibitemOpen
  \bibfield  {author} {\bibinfo {author} {\bibfnamefont {B.}~\bibnamefont
  {Vlastakis}}, \bibinfo {author} {\bibfnamefont {G.}~\bibnamefont
  {Kirchmair}}, \bibinfo {author} {\bibfnamefont {Z.}~\bibnamefont {Leghtas}},
  \bibinfo {author} {\bibfnamefont {S.~E.}\ \bibnamefont {Nigg}}, \bibinfo
  {author} {\bibfnamefont {L.}~\bibnamefont {Frunzio}}, \bibinfo {author}
  {\bibfnamefont {S.~M.}\ \bibnamefont {Girvin}}, \bibinfo {author}
  {\bibfnamefont {M.}~\bibnamefont {Mirrahimi}}, \bibinfo {author}
  {\bibfnamefont {M.}~\bibnamefont {Devoret}}, \ and\ \bibinfo {author}
  {\bibfnamefont {R.~J.}\ \bibnamefont {Schoelkopf}},\ }\href {\doibase
  10.1126/science.1243289} {\bibfield  {journal} {\bibinfo  {journal}
  {Science}\ }\textbf {\bibinfo {volume} {342}},\ \bibinfo {pages} {607}
  (\bibinfo {year} {2013})}\BibitemShut {NoStop}%
\bibitem [{\citenamefont {Albert}\ \emph {et~al.}(2016)\citenamefont {Albert},
  \citenamefont {Shu}, \citenamefont {Krastanov}, \citenamefont {Shen},
  \citenamefont {Liu}, \citenamefont {Yang}, \citenamefont {Schoelkopf},
  \citenamefont {Mirrahimi}, \citenamefont {Devoret},\ and\ \citenamefont
  {Jiang}}]{Albert2015}%
  \BibitemOpen
  \bibfield  {author} {\bibinfo {author} {\bibfnamefont {V.~V.}\ \bibnamefont
  {Albert}}, \bibinfo {author} {\bibfnamefont {C.}~\bibnamefont {Shu}},
  \bibinfo {author} {\bibfnamefont {S.}~\bibnamefont {Krastanov}}, \bibinfo
  {author} {\bibfnamefont {C.}~\bibnamefont {Shen}}, \bibinfo {author}
  {\bibfnamefont {R.-B.}\ \bibnamefont {Liu}}, \bibinfo {author} {\bibfnamefont
  {Z.-B.}\ \bibnamefont {Yang}}, \bibinfo {author} {\bibfnamefont {R.~J.}\
  \bibnamefont {Schoelkopf}}, \bibinfo {author} {\bibfnamefont
  {M.}~\bibnamefont {Mirrahimi}}, \bibinfo {author} {\bibfnamefont {M.~H.}\
  \bibnamefont {Devoret}}, \ and\ \bibinfo {author} {\bibfnamefont
  {L.}~\bibnamefont {Jiang}},\ }\href {\doibase 10.1103/PhysRevLett.116.140502}
  {\bibfield  {journal} {\bibinfo  {journal} {Phys. Rev. Lett.}\ }\textbf
  {\bibinfo {volume} {116}},\ \bibinfo {pages} {140502} (\bibinfo {year}
  {2016})}\BibitemShut {NoStop}%
\bibitem [{\citenamefont {Heeres}\ \emph {et~al.}(2015)\citenamefont {Heeres},
  \citenamefont {Vlastakis}, \citenamefont {Holland}, \citenamefont
  {Krastanov}, \citenamefont {Albert}, \citenamefont {Frunzio}, \citenamefont
  {Jiang},\ and\ \citenamefont {Schoelkopf}}]{Heeres2015}%
  \BibitemOpen
  \bibfield  {author} {\bibinfo {author} {\bibfnamefont {R.~W.}\ \bibnamefont
  {Heeres}}, \bibinfo {author} {\bibfnamefont {B.}~\bibnamefont {Vlastakis}},
  \bibinfo {author} {\bibfnamefont {E.}~\bibnamefont {Holland}}, \bibinfo
  {author} {\bibfnamefont {S.}~\bibnamefont {Krastanov}}, \bibinfo {author}
  {\bibfnamefont {V.~V.}\ \bibnamefont {Albert}}, \bibinfo {author}
  {\bibfnamefont {L.}~\bibnamefont {Frunzio}}, \bibinfo {author} {\bibfnamefont
  {L.}~\bibnamefont {Jiang}}, \ and\ \bibinfo {author} {\bibfnamefont {R.~J.}\
  \bibnamefont {Schoelkopf}},\ }\href {\doibase
  http://dx.doi.org/10.1103/PhysRevLett.115.137002} {\bibfield  {journal}
  {\bibinfo  {journal} {Phys. Rev. Lett.}\ }\textbf {\bibinfo {volume} {115}},\
  \bibinfo {pages} {137002} (\bibinfo {year} {2015})}\BibitemShut {NoStop}%
	\bibitem [{vBp()}]{vBp_suppl}%
  \BibitemOpen
  \href@noop {} {\bibinfo  {journal} {See accompanying supplementary materials
  for details}\ }\BibitemShut {NoStop}%
\end{thebibliography}

%

\clearpage

\onecolumngrid
\appendix

%% Special for the Suppl
\newcommand*{\fogiStark}{\Delta^{(\Omega)}_{f0g1}}
\newcommand*{\dispStark}{\Delta^{(n)}}
\newcommand*{\phsh}{\varphi_s}
\newcommand*{\phappl}{\varphi_0}
\newcommand*{\inPsi}{\Psi_n^{(i)}}

%% Suppl Mat numbering
\setcounter{equation}{0}
\setcounter{figure}{0}
\renewcommand{\theequation}{S\arabic{equation}}
\renewcommand{\thefigure}{S\arabic{figure}}
\renewcommand{\figurename}{fig.}

\section*{Supplementary Materials}

\subsection*{Characterization of the tunable coupling}

We characterize the tunable coupling between states $\ket{f,n}$ and $\ket{g,n+1}$ [24] by Rabi spectroscopy. We use the first cavity mode for the dispersive readout of the transmon states. We observe Rabi oscillations between the states $\ket{g}$ and $\ket{f}$ by varying the duration $\tau$ of a square Gaussian pulse of fixed amplitude $\Omega$ and a $3~\rm{ns}$ rise time. Sample oscillations are shown in fig.~S1A for different detunings $\Delta$.
%We determine the resonant frequency $\omega_d$ and the resonant coupling strength $g$ from the dependence of the Rabi frequency $\Omega_R$ on $\Delta$ [Fig. S1(B)].
We record the dependence of the Rabi frequency $\Omega_R$ on $\Delta$ (fig. S1B)
and determine the resonant frequency $\omega_d$ and the resonant coupling strength $g$ from the minimum of $\Omega_R$. We repeat this measurement for different pulse amplitudes $\Omega$. We find that the coupling strength $g$ increases linearly with $\Omega$ (fig.~S2A) and that the resonant frequency $\omega_d$
%$\omega_d(\Omega)=\omega_d(0)+\Delta_{f0g1}(\Omega)$
is significantly Stark-shifted by the drive (fig.~S2B). This Stark shift $\fogiStark$ is quadratic in $\Omega$ in the relevant parameter range.
%For a typical coupling strength used in the experiment, $g/2\pi\approx 4~\rm{MHz}$, the corresponding Stark shift is $\fogiStark/2\pi\approx30~\rm{MHz}$.

\subsection*{Characterization at higher photon numbers}

To prepare a given Fock state in the cavity, we iteratively apply a sequence of $\pi$ pulses to the $\trans{g}{e}$, $\trans{e}{f}$, and $\trans{f,n}{g,n+1}$ transitions. The resonant frequencies of these transitions are Stark shifted due to the photon-number-dependent dispersive shifts $\dispStark_{ge}$, $\dispStark_{ef}$, and $\dispStark_{f,n;g,n+1}$. We measure $\dispStark_{ge}$ and $\dispStark_{ef}$ by Ramsey spectroscopy and $\dispStark_{f,n;g,n+1}$ with the technique described in the previous section (fig.~S3A). We find that $\dispStark_{f,n;g,n+1} \approx \dispStark_{ge}+\dispStark_{ef}$. 
Finally, we measure the coupling $g_n$ between the states $\ket{f,n}$ and $\ket{g,n+1}$. In fig.~S3B we plot $g_n^2$ versus the photon number $n$ and for two different amplitudes of the coupling drive. We find that $g_n^2 \propto n+1$, in agreement with the theory prediction [24].

\subsection*{Geometric phases acquired by the eigenstates $\eigens$}

When the tunable coupling is on, the Hamiltonian in the $\{ \ket{f,n}, \ket{g,n+1} \}$ subspace, in a frame rotating at the drive frequency and after the rotating-wave approximation, is
\be
\mathcal H_n = \twobytwo{-\Delta/2}{g\sqrt{n+1}e^{i \varphi}}{g\sqrt{n+1}e^{-i \varphi}}{\Delta/2}  \label{eq:Hn}
\ee
The instantaneous eigenstates of $\mathcal H_n$ can be written as
\begin{align*}
\ket{\Psi_n^-(\varphi)} &= -\sin (\mixngl/2) e^{i \varphi} \ket{g,n+1} + \cos (\mixngl/2) \ket{f,n} 
\\
\ket{\Psi_n^+(\varphi)} &= \cos (\mixngl/2) e^{i \varphi} \ket{g,n+1} +\sin (\mixngl/2) \ket{f,n} 
\end{align*}
where the mixing angle $\mixngl$ is defined by $\tan(\mixngl)= 2\coup\sqrt{\nph+1}/\dtn$.
If $\mixngl$ is kept constant while $\varphi$ is slowly varied between $0$ and $\dphi$ during a time $\tau$, the corresponding geometric phase acquired by $\eigens$ is given by
\be
\bar\gamma_n^\pm(\dphi) = i\int_0^{\dphi} d\varphi \braket{\Psi_n^\pm}{\frac{d}{d\varphi}\Psi_n^\pm} = -\left( 1 \pm \cos \mixngl \right) \frac{\dphi}{2}  \label{eq:geomph}
\ee
At the same time, each eigenstate also acquire a dynamic phase
\be
\xi_n^\pm = \pm \frac{\tau}{2} \sqrt{\Delta^2+4g^2(n+1)} \label{eq:dynph}
\ee

\subsection*{Geometric phase estimation based on open loops}

The protocol described in Fig.~4A allows us to determine the geometric phase acquired between the two eigenstates $\eigmn$ and $\eigpl$ by utilizing a superposition of them for interferometry. In contrast to other interferometric schemes, the geometric phase is continuously tracked as it is acquired along the loop. For simplicity, we here illustrate a reduced sequence in which a single coupling pulse is used. We discuss the two-pulse, spin-echo-type sequence of Fig.~4A in the next section.

After preparing the initial state $\ket{\inPsi}$, we quickly turn on the coupling. Then we vary the coupling phase $\varphi$ between $0$ and $\dphi$ during a time $\tau$. Finally, we turn off the coupling and measure the state of the transmon. Neglecting decoherence effects, the probability of finding the system in the initial state $\ket{\inPsi}$ at the end of the sequence is given by
\be
\begin{split}
P &= \left|\sum_{\alpha=+,-}\braket{\inPsi}{\Psi_n^{\alpha}(\delta\varphi)}e^{i \xi_n^\alpha +i\gamma_n^\alpha} \braket{\Psi_n^{\alpha}(0)}{\inPsi}\right|^2 \\
&= (|A_+|-|A_-|)^2 + 4 |A_+| |A_-| \cos^2(\chi_+-\chi_-) 
\end{split} \label{eq:P_open_gen}
\ee
with
\bes
\begin{split}
A_\alpha &= \braket{\inPsi}{\Psi_n^{\alpha}(\delta\varphi)}\braket{\Psi_n^{\alpha}(0)}{\inPsi} \\
\chi_\alpha &= \xi_n^\alpha+ \gamma_n^\alpha+ \arg\left[\braket{\Psi_n^{\alpha}(0)}{\inPsi}\right] - \arg \left[ \braket{\Psi_n^{\alpha}(\delta\varphi)}{\inPsi} \right] 
\end{split}
\ees
For our chosen initial state $\ket{\inPsi}=\ket{f,n}$, the ``$\arg$'' terms in the expression for $\chi_\alpha$ vanish identically
%one has $A_-=\cos^2(\mixngl/2)$, $A_+=\sin^2(\mixngl/2)$, and $\chi_\alpha &= \xi_n^\alpha+ \gamma_n^\alpha$.
and the probability simplifies to
\be
P = \cos^2(\mixngl) + \sin^2 (\mixngl) \cos^2\left[\frac12(\xi_n^+ - \xi_n^- + \gamma_n^+ - \gamma_n^-)\right]  \label{eq:P_open_fn}
\ee
Equation \ref{eq:P_open_fn} describes an interference pattern with visibility $\sin^2 (\mixngl)$. The interference phase consists of a dynamic contribution $\delta\xi = \xi_n^+ - \xi_n^-$ and a geometric contribution $\delta\gamma=\gamma_n^+ - \gamma_n^-$. While $\delta\xi$ does not depend on $\dphi$ (see eq.~\ref{eq:dynph}), $\delta\gamma$ is linearly proportional to $\dphi$, according to eq.~\eqref{eq:geomph}. The geometric contribution $\delta\gamma$ can be visualized on the Bloch sphere as in fig.~S4, where the open paths described by the two eigenstates $\eigmn$ and $\eigpl$  (solid lines) are connected to the initial state $\ket{f,n}$ by means of geodesic paths (dashed lines), according to the standard prescription [28]. This procedure defines two solid angles (indicated in blue and red) whose difference (orange) is proportional to $\delta\gamma$.

The directly proportional relationship between the measured geometric phase $\delta\gamma$ and the coupling-phase variation $\dphi$ is key to our phase-extraction method. As clear from the geometric argument illustrated above, this is due to the following two reasons: (i) the symmetry of the path described by the system, which spans a circle of latitude on the Bloch sphere, and (ii) the choice of the north pole of the sphere, $\ket{f,n}$, as the initial state.

\subsection*{Dynamic phase cancellation}

In order to remove the dynamic-phase contribution to the interference pattern in eq.~\ref{eq:P_open_fn}, we repeat the phase-varying pulse twice (see Fig.~4A). During the second pulse, the effective Hamiltonian is
\bes
\mathcal H_n^{(2)} = e^{-i \sigma_y \frac\pi2} \mathcal H_n e^{i \sigma_y \frac\pi2} 
%=
%\twobytwo{\Delta/2}{-g\sqrt{n+1}e^{-i \varphi}}{-g\sqrt{n+1}e^{i \varphi}}{-\Delta/2}
\ees
which is obtained from $\mathcal H_n$ by the substitutions $\Delta \to -\Delta$ and $\varphi \to \pi - \varphi$. This choice effectively realizes a spin echo, eliminating the dynamic-phase contribution and doubling the acquired geometric phase.

\subsection*{Insensitivity to small deviations from the resonant frequency}

In the geometric-phase estimation protocol described in Fig.~4A, an erroneous estimation of the resonant frequency $\omega_d$ for the $\trans{f,n}{g,n+1}$ transition produces a phase shift of the patterns of Fig.~4B, due to residual dynamic-phase contributions. However, the frequency of the oscillations and, hence, the extracted geometric phase remain unaffected. In fact,
%assume that the estimated resonant frequency is off by a small amount $\delta\omega_d$, so that $\Delta \to \Delta - \delta\omega_d$.
as the phase of the coupling is varied by an amount $\dphi$ during a time $\tau$, a small, additional detuning $\delta\omega_d$ generates an additional phase shift $\delta\omega_d \tau$.
%To see this, it is enough to consider the Hamiltonian \eqref{eq:Hn} and apply a time-dependent rotation $e^{i \delta\omega_d \sigma_z/2}$, where $\sigma_z$ is the third Pauli matrix. This property extends to the full spin-echo protocol, with the corresponding phase shifts summing up.

In the experiment of Fig.~4, the calibration of $\omega_d$ at arbitrary detunings is problematic due to the fact that the microwave tone used to drive the transition induces a strong ac Stark shift on the transition itself (see fig.~S2B). As a result, uncertainties in the drive amplitude (due to, e.~g., imperfect mixer calibration or frequency-dependent attenuation in the lines) directly translate into frequency errors. We ascribe the different phase shifts observed in the traces of Fig.~4B to this effect. The maximum phase shift in the series is $0.89~\rm{rad}$ (when $\Delta/2\pi=10~\rm{MHz}$), corresponding to a frequency error of about $90~\rm{kHz}$ during each half of the spin-echo protocol.

\subsection*{Phase calibration of spin-echo coupling pulses}

As all transition frequencies are Stark-shifted when the tunable coupling is on, particular care must be taken to correct for phase shifts. During the pulse sequence of Fig.~4A, a phase shift $\phsh$ is accumulated in the reference frame of the drive as the coupling is switched off in the middle of the sequence.
This phase shift can be calibrated by running a pulse sequence consisting of two resonant coupling pulses ($\Delta=0$) of identical length $\tau$ and a varying phase shift $\phappl$ between them.
%and $\varphi_2$ during the two halves of the protocol.
%The phase shift $\dphi_{\rm sh}$ is then determined by the condition that the final state should be closest to the initial state when $\varphi_2-\varphi_1-\dphi_{\rm sh}=\pi$.
For a generic value of $\phappl$, the system undergoes Rabi oscillations as a function of the pulse length $\tau$. However, when the condition $\phappl-\phsh=\pi$ is met, the amplitude of the oscillations approaches a minimum as the evolution is closest to a perfect spin echo. We use this condition to experimentally measure $\phsh$ and compare its value to an independent estimate based on the measured Stark shift and the time separation between the two pulses. We find a good agreement between the two estimates, their difference corresponding to an uncertainty of about $0.2~\rm{ns}$ on the separation between the pulses.

\clearpage
\subsection*{Supplementary Figures}

\begin{figure*}[h]
\centering
	\includegraphics{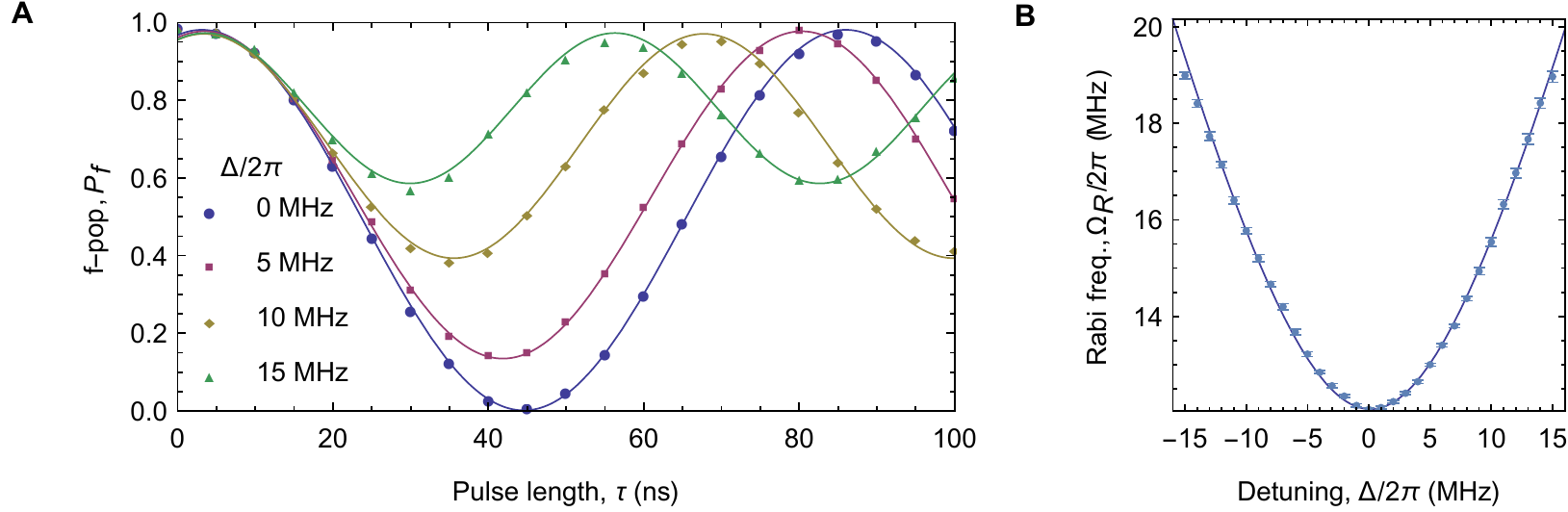}
	\caption{Rabi spectroscopy of the tunable coupling.
\pA $f$-state population $P_f$ versus pulse length $\tau$ for different detunings $\Delta$ (symbols: data; solid lines: sine fits).
\pB Rabi frequency $\Omega_R$ versus $\Delta$, as obtained from the measurements in (A) (points: data; solid line: best-fit of the model expression $\Omega_R = \sqrt{\Delta^2+4g^2}$).}
	\label{fig:fogi_Rabi}
\end{figure*}

%\clearpage
\begin{figure}
\centering
	\includegraphics{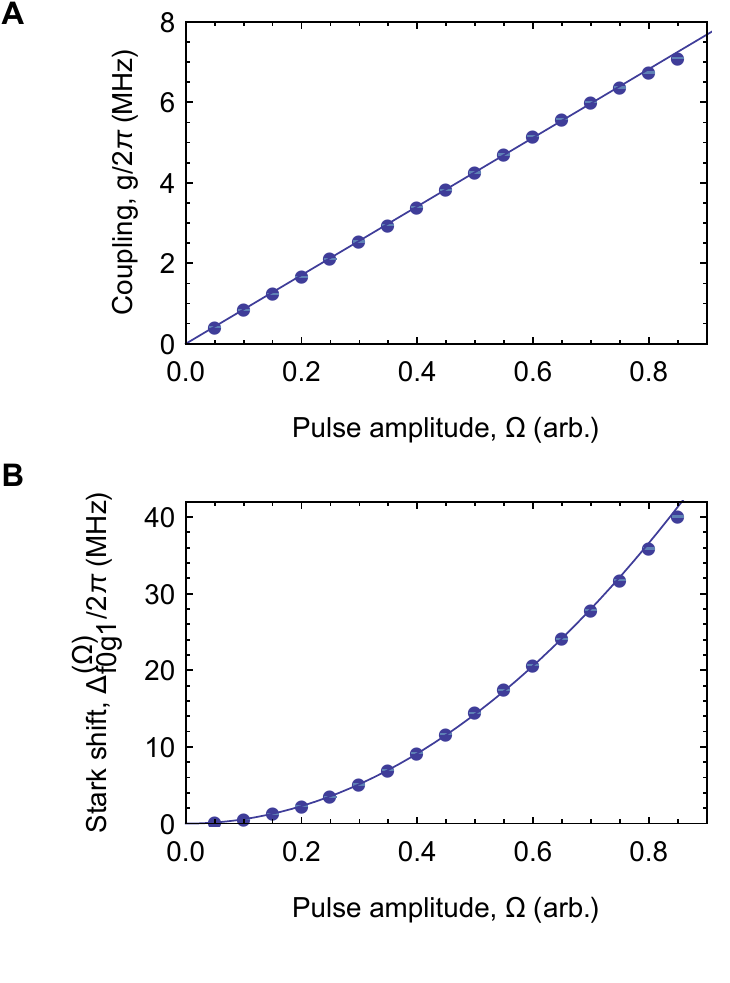}
	\caption{Calibration of strength and resonant frequency of the tunable coupling.
\pA Coupling strength $g$ and \pB Stark shift $\fogiStark$ versus pulse amplitude $\Omega$, as obtained from Rabi spectroscopy. The solid lines are linear (A) and quadratic (B) fits to the data.}
	\label{fig:ampl_calib}
\end{figure}

%\clearpage
\begin{figure}
\centering
	\includegraphics{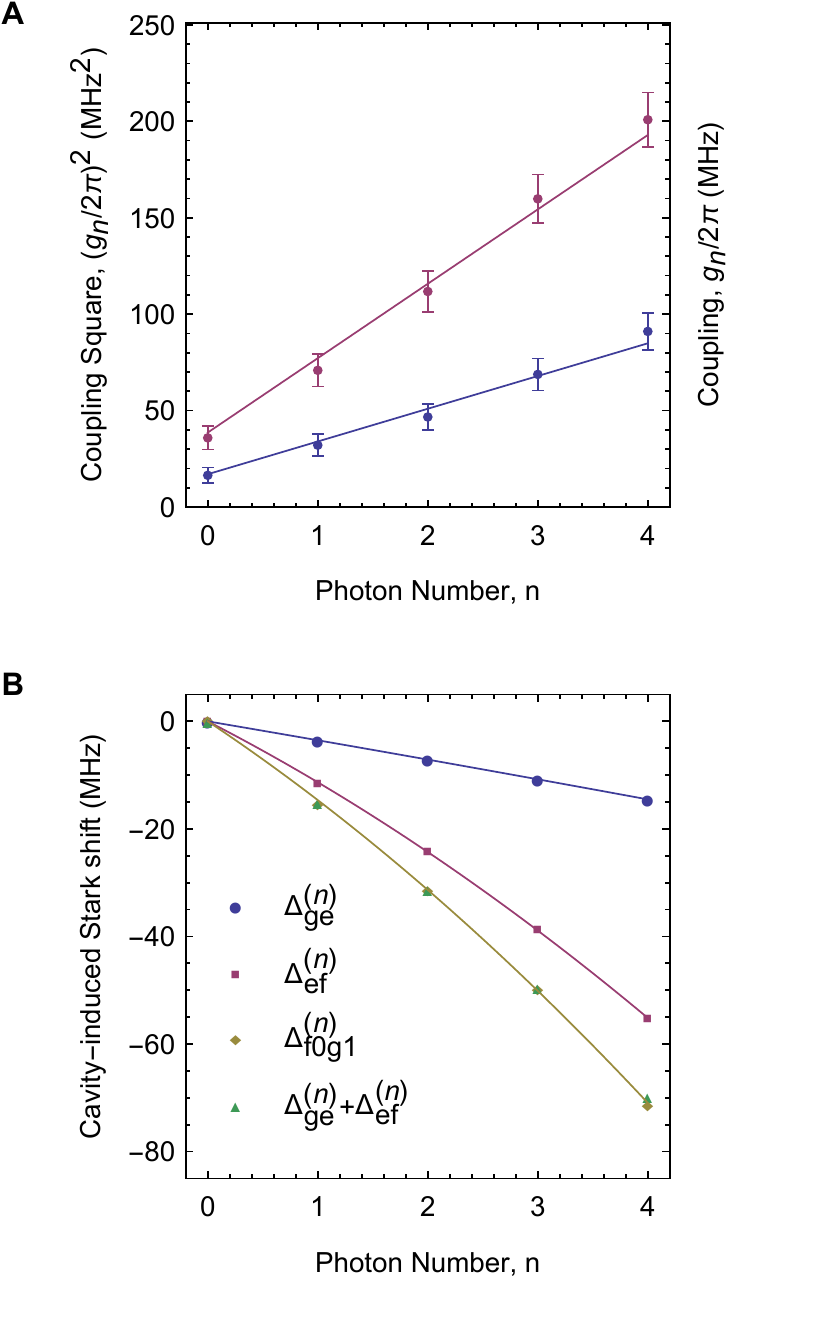}
		\caption{Calibration at higher photon numbers. 
\pA Square of the coupling $g_n$ between the states $\ket{f,n}$ and $\ket{g,n+1}$ versus photon number $n$, as measured by Rabi spectroscopy, for two different field amplitudes $\Omega$ (blue and red dots). The solid lines are linear fits of the Jaynes-Cummings expression $g_n^2 = g^2(n+1)$ to each data set. From the two fits we extract $g/2\pi=(4.12 \pm 0.06)~\rm{MHz}$ and $g/2\pi=(6.21 \pm 0.08)~\rm{MHz}$, respectively.
\pB Measured photon-number-dependent Stark shifts $\dispStark_{ge}$, $\dispStark_{ef}$ and $\dispStark_{f,n;g,n+1}$ (symbols). The combined shift $\dispStark_{ge}+\dispStark_{ef}$ is plotted for comparison. The solid lines are second-order polynomial fits to the data. }
	\label{fig:photon_calib}
\end{figure}

%\clearpage
\begin{figure}
\centering
	\includegraphics{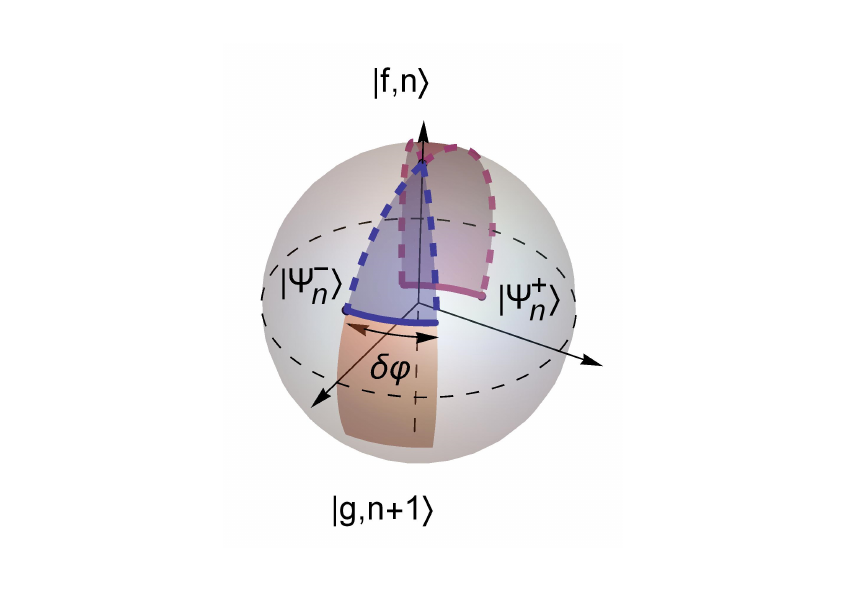}
		\caption{Geometric interpretation of the open-loop protocol. 
As the phase of the coupling is changed by an amount $\dphi$, the two eigenstates $\eigens$ describe open paths on the Bloch sphere (solid lines, blue and red, respectively). To compute the open-loop geometric phase acquired by the state $\ket{f,n}$, we add geodesic paths connecting the the ends of each open path to $\ket{f,n}$ (dashed lines). This construction defines two solid angles (indicated in blue and red), whose difference (orange) is proportional to the geometric phase acquired by $\ket{f,n}$ and measured in our experiment.
}
	\label{fig:open_loop}
\end{figure}

\end{document}